\documentclass{JHEP3}
\usepackage{amssymb}

\usepackage{amsmath,amssymb,graphics}
\usepackage{epsfig,multicol}
\usepackage{graphicx}
\usepackage{epsfig,subfigure}

\newcommand {\nn}    {\nonumber}

\title{Localization of Matters on Anti-de Sitter Thick Branes}

\author{Yu-Xiao Liu,
        Heng Guo\footnote{Corresponding author.},
        Chun-E Fu,
        Ji-Rong Ren \\
  Institute of Theoretical Physics,
  Lanzhou University, Lanzhou 730000,
  People's Republic of China\\
  E-mail: \email{liuyx@lzu.edu.cn}, \email{guoh2009@lzu.cn},
          \email{fuche08@lzu.cn}, \email{renjr@lzu.edu.cn}}

\abstract{ By presenting the mass-independent potentials of the
Kaluza-Klein (KK) modes in the corresponding Schr\"{o}dinger
equations, we investigate the localization and mass spectra of
various bulk matter fields on an AdS thick brane. For a spin 0 scalar
$\Phi$ coupled with itself and the domain-wall-forming field $\phi$
via a coupling potential
 $V=(\lambda \phi^{2}-u^{2})\Phi^{2}+\tau\Phi^{4}$,
the localization and spectrum are decided by a critical coupling
constant $\lambda_{0}$. When $\lambda>\lambda_0$, the potential of
the scalar KK modes in the corresponding Schr\"{o}dinger equation
tends to infinite when far away from the brane, which results in
that there exist infinite discrete scalar bound KK states, and the
massless modes could be trapped on the AdS brane by fine-tuning of
parameters. When $\lambda<\lambda_0$, the potential of the scalar KK
modes tends to negative infinite when far away from the brane,
hence there does not exist any scalar bound KK state. For a spin 1
vector, the situation is same like the scalar with a coupling
constant $\lambda>\lambda_0$, but the zero mode can not be localized on the brane.
For a spin 1/2 fermion, we introduce
the usual Yukawa coupling $\eta\bar{\Psi}\phi\Psi$, and find that
the localization of the fermion is decided by a critical coupling
constant $\eta_0$. For $\eta > \eta_0$, the four-dimensional
massless left chiral fermion and massive Dirac fermions consisted of
the pairs of coupled left-hand and right-hand KK modes could be
localized on the AdS brane, and the massive Dirac fermions have a
set of discrete mass spectrum. While for the case $0<\eta <
\eta_{0}$, no four-dimensional Dirac fermion can be localized on the
AdS brane.}


\begin{document}

\section{Introduction}

The proposal that our observed four-dimensional universe as a
sub-manifold is embedding in a higher dimensional space, has
received a considerable attention. Branes naturally appear in
string/M-theory context and provide a novel mode for discussing
phenomenological and cosmological issues related to extra
dimensions. The idea that extra dimensions may not be compact
\cite{RubakovPLB1983136,VisserPLB1985,Randjbar-DaemiPLB1986,rs,Lykken}
or large \cite{AntoniadisPLB1990,ADD} can supply new insights for
solving gauge hierarch problem \cite{ADD} and cosmological
constant problem
\cite{RubakovPLB1983136,Randjbar-DaemiPLB1986,CosmConst}. The
framework of brane scenarios is that gravity is free to propagate
in all dimensions, however all matter fields are confined to a
3-brane with no contradiction with present gravitational
experiments \cite{RubakovPLB1983136,VisserPLB1985,ADD}. In
Randall-Sundrum (RS) brane model \cite{rs}, the internal manifold
does not need to be compactified to the Planck scale any more, it
can be large, or even infinite non-compact. But in this model, the
branes are very idealized because of their thickness is neglected.
It is also widely considered that the most fundamental theory
would have a minimal length scale.

Since thick brane scenarios based on gravity coupled to scalars have
been constructed, more and more authors have investigated the thick
brane scenario in higher dimensional space-time
\cite{dewolfe,gremm,Csaki,CamposPRL2002,WangPRD2002,varios,ThickBrane,
Liu0907.1952,VolkasPRD2008,VolkasJHEP09,DaviesPRD07}. In these
models, there is an interesting feature that we can obtain branes
naturally without introducing them by hand in the action of the
theory \cite{dewolfe}. In this scenario the scalar fields do not
play the role of bulk fields but provide the ``material" from which
the thick branes are made of. For a comprehensive review on thick
branes please see Ref. \cite{ThickBraneReview}.

In brane world scenarios, there is an important problem: whether
various bulk fields could be localized on the brane by a natural
mechanism. Generally, massless scalar fields \cite{BajcPLB2000} and
graviton \cite{rs} can be localized on brane of different types.
Vector fields can be localized on the RS brane in some
higher-dimensional cases \cite{OdaPLB2000113} or on the thick $dS$
brane and Weyl thick brane \cite{Liu0708}. It is important that
whether the fermions could be localized on the thick branes. Without
introducing the scalar-fermion coupling, fermions do not have
normalizable zero modes in five and six dimensions
\cite{BajcPLB2000,OdaPLB2000113,Liu0708,NonLocalizedFermion,
VolkasJHEP0704,IchinosePRD2002,Ringeval,GherghettaPRL2000,
Neupane,RandjbarPLB2000,KoleyCQG2005,DubovskyPRD2000,0803.1458,
LiuJHEP2007,0901.3543,Liu0907.0910,Koley2009,0812.2638,Liu0911}. In
some cases, with scalar-fermion coupling, one can obtain a single
bound state and a continuous gapless spectrum of massive fermion
Kaluza--Klein (KK) states \cite{Liu0708,ThickBraneWeyl}. In some
other brane models, one can obtain finite discrete KK states (mass
gap) and a continuous gapless spectrum starting at a positive $m^2$
\cite{ThickBrane4,LiuJCAP2009,Liu0803}.

In Ref. \cite{LiuJCAP2009}, the authors have investigated the
localization and mass spectra of various matter fields on the
symmetric and asymmetric dS thick branes. For scalar and vector
fields, the potentials of the KK modes in the corresponding
Schr\"{o}dinger equations are the modified P\"{o}schl-Teller
potentials. In this case, there exist finite mass gap and a series
of continuous spectrum. For fermions, when one introduces the
usual Yukawa coupling, there exists no mass gap but a continuous
gapless spectrum of KK states. While in Ref. \cite{Liu0803}, the
same problem of various matter fields on a family of thick brane
configurations in a pure geometric 5-dimensional spacetime was
studied. It was shown that, for a class of brane configurations,
there exists a continuum gapless spectrum of KK modes with any
$m^2>0$ for scalars, vectors and ones of left chiral and right
chiral fermions. However, for a special of brane configuration,
there exist mass gap and a series of continuous spectrum starting
at positive $m^2$ for scalars, vectors and fermions.

Recently, ``Locally localized gravity" solutions with AdS$_4$ and
dS$_4$ brane cosmology has been presented in Ref.
\cite{VolkasJHEP0704}. In the case of a brane with AdS$_4$
cosmology, the warp factor grows exponentially when far from the
brane. However, when closes to the brane, the metric behaves like
the decreasing warped metric associated with a Minkowski brane, and
it is this local behavior that is responsible for confining gravity
to the brane \cite{VolkasJHEP0704,KobayashiPRD2002,RandallJHEP2001}.
Confinement of the fermion zero mode on the AdS$_4$ brane has been
discussed in Ref. \cite{VolkasJHEP0704}, and the localization
conditions for left-handed and right-handed fermion zero modes were
obtained. In this paper, in order to reveal the rich structures of
the AdS$_4$ brane from other point of view, we would like to
investigate further the localization problem of various spin massive
mater fields (scalars, vectors and  fermions) on the brane. We find
the results especially for scalars and fermions are very nontrivial
and have not been reported in the literature: For a spin-0 scalar
$\Phi$ coupled with itself and the background scalar $\phi$ via the
coupling potential $(\lambda \phi^{2}-u^{2})\Phi^{2}+\tau\Phi^{4}$,
we find that the localization characteristic is related closely to a
critical coupling constant $\lambda_{0}$. When $\lambda >
\lambda_{0}$, there exist a series of discrete bound KK modes and
the zero mode could be trapped on the AdS thick brane by fine-tuning
of parameters. However, when $\lambda < \lambda_{0}$, there does not
exist any bound KK modes. For a spin-1 vector, there exist infinite
massive bound KK modes, but the massless mode could not be trapped
on the AdS brane. For a fermion coupled with the background scalar
via the usual Yukawa coupling $\eta\bar{\Psi}\phi\Psi$, in order to
get a series of four-dimensional massive Dirac fermions and a
massless left chiral fermion, the coupling constant $\eta$ should
larger than a critical coupling constant $\eta_0$.

The organization of the paper is as follows: In Sec. \ref{SecModel},
we first review the AdS thick brane in 5-dimensional space-time.
Then, in Sec. \ref{SecLocalize}, we study the localization and mass
spectra of various bulk fields on the AdS thick brane by presenting
the potentials of the corresponding Schr\"{o}dinger problem. For
scalar fields, we introduce the Higgs coupling and discuss the
spectra of scalars in detail. For vector fields, we obtain infinite
bound states and infinite mass gaps. For spin 1/2 fermions, we
introduce the usual Yukawa coupling and discuss the spectra of
fermions in detail. Finally, our conclusion is given in Sec.
\ref{SecConclusion}.

\section{Review of anti-de Sitter thick brane}
\label{SecModel}

Let us consider thick branes arising from a real scalar field
$\phi$ with a scalar potential $V(\phi)$. The action for such a
system can be expressed as
 \begin{equation}
S = \int d^5 x \sqrt{-g}\left [ \frac{1}{2\kappa_5^2} R-\frac{1}{2}
g^{MN}\partial_M \phi
\partial_N \phi - V(\phi) \right ],
\label{action}
\end{equation}
where $R$ is the scalar curvature and $\kappa_5^2=8 \pi G_5$ with
$G_5$ the 5-dimensional Newton constant. Here we set $\kappa_5=1$.
The line-element for a 5-dimensional spacetime is assumed as
\begin{eqnarray}\label{linee}
 ds^2 = g_{MN}dx^{M}dx^{N}
 =\text{e}^{2A(y)}\hat{g}_{\mu\nu}(x)dx^\mu dx^\nu + dy^2,
\end{eqnarray}
where $\text{e}^{2A(y)}$ is the warp factor and $y$ stands for the
extra coordinate. We suppose that $\hat{g}_{\mu\nu}$ is some
general 4-dimensional metric. The scalar field is considered to be
a function of $y$ only, i.e., $\phi=\phi(y)$. In the model, the
potential could provide a thick brane realization, and the soliton
configuration of the scalar field dynamically generates the domain
wall with warped geometry. The field equations generated from the
action (\ref{action}) with the ansatz (\ref{linee}) reduce to the
following coupled nonlinear differential equations
\begin{eqnarray}
\phi'^{2} & = & -3A''- \gamma e^{-2A}, \label{phi'2} \\
V(\phi) & = &  \frac{1}{2}\phi'^{2} - 6A'^{2} + 2\gamma e^{-2A} , \label{Vy} \\
\frac{dV(\phi)}{d\phi} &  = & \phi'' + 4A'\phi', \label{dVphi}
\end{eqnarray}
where the prime denotes derivative with respect to $y$ and
$\gamma$ is some constant such that $\hat{G}_{\mu\nu}=\gamma
\hat{g}_{\mu\nu}$. In the case of an AdS brane cosmology, $\gamma$
is negative, while for a dS brane cosmology we have $\gamma>0$.
Minkowski space corresponds to the case $\gamma=0$.

We consider a trial warp factor of the form \cite{VolkasJHEP0704},
\begin{eqnarray}\label{eA}
 e^{A(y)}= a \cosh(c y)+ b \; \textrm{sech}(c y),
\end{eqnarray}
where $a$ and $b$ are dimensionless constants and
 $c= r \sqrt{\frac{|\gamma|}{3}}$ with $r$ a positive constant.
Then Eq. (\ref{phi'2}) becomes
\begin{eqnarray}\label{phi'22}
 \phi'^{2} = \frac{-|\gamma|}{\big[a \cosh(c y)
       + b \;\textrm{sech}(c y)\big]^2}
     \bigg\{ \frac{\gamma}{|\gamma|}
     + r^2 \bigg[a(a + 4 b) - 4 a b \;\textrm{sech}^2(c y)
     - b^2 \textrm{sech}^4 (c y) \bigg] \bigg\}.~~~
\end{eqnarray}
If the r.h.s. of the above equation is always positive and to
facilitate an analytic solution, we should impose the relation
\begin{eqnarray}\label{gamma}
 - \frac{\gamma}{|\gamma|} = r^2 \left( 4 a^2 + a( a + 4 b)\right).
\end{eqnarray}
Then Eq. (\ref{phi'22}) reduces to
\begin{eqnarray}
 \phi'^2  =  |\gamma| r^2 \textrm{sech}^2 (c y)
    \left( \frac{2 a + b\; \textrm{sech}^2 (c y)}{a + b\;
     \textrm{sech}^2(c y)} \right)^2,
\end{eqnarray}
which can be solved as \cite{VolkasJHEP0704}
\begin{eqnarray}
\phi(y) &=& \pm \sqrt{3}\bigg[ \arctan \sinh (c y) \
             + \sqrt{\frac{a}{a+b}} \arctan
               \left( \sqrt{\frac{a}{a+b}} \sinh (c y) \right) \bigg].
\end{eqnarray}
The potential $V$ can easily be expressed as a function of $y$:
\begin{eqnarray}
 V(y) =  \frac{r^2 |\gamma|}{\big[ a + b\; \textrm{sech}^2 ( c y) \big]^2}
      && \bigg[ - 2 a (3 a + 2 b) \textrm{sech}^2 ( c y)
            - 2 b (a + b) \textrm{sech}^4 (c y)  \nonumber \\
      &&~ + \frac{5}{2} b^2 \textrm{sech}^6 (c y)
           - 2 a^2 \bigg].~~~~~
\end{eqnarray}
Generally, $V$ can not be written analytically in term of standard
function of $\phi$, however $V$ can always be expressed
numerically as a well-defined function of $\phi$.

In this paper we would like to investigate the AdS brane
cosmology, so $\gamma$ is negative, and by Eq. (\ref{gamma}), this
is equivalent to requiring
\begin{eqnarray}
5 a^2 + 4ab > 0.
\end{eqnarray}

\section{Localization of various matters on an AdS thick brane}
\label{SecLocalize}

In this section, we will investigate whether various bulk mater
fields such as spin-0 scalars, spin-1 vectors and spin-1/2
fermions can be localized on the AdS thick brane by means of the
gravitational interaction. Certainly, we have implicitly assumed
that various bulk mater fields considered below make little
contribution to the bulk energy so that the solutions given in
previous section remain valid even in the presence of bulk fields.
We will also discuss the spectra of various mater fields on the
AdS thick brane by presenting the potential of the corresponding
Schr\"{o}dinger equation for KK modes of various matter fields.

In order to get mass-independent potentials, we will follow Ref.
\cite{rs} and change the metric given in (\ref{linee}) to
following one
\begin{equation}
\label{conflinee2}
 ds^2=\text{e}^{2A(z)}\left(\hat{g}_{\mu\nu}dx^\mu dx^\nu+dz^2\right)
\end{equation}
by performing the coordinate transformation
\begin{equation}\label{transformation}
dz=\text{e}^{-A(y)}dy.
\end{equation}
Then, we can obtain the following expression
\begin{eqnarray}
z(y)=\int e^{-A(y)}d y
    =\sqrt{\frac{1}{a(a+b)}}
    \frac{1}{c}\arctan\left[\sqrt{\frac{a}{a+b}}\sinh(c y)\right].
    \label{z_y}
\end{eqnarray}
From this expression, it can be seen that $z\rightarrow \pm
\frac{\pi}{2c}\sqrt{\frac{1}{a(a+b)}}$, when $y\rightarrow
\pm\infty$, so the range of $z$ is $-z_{max}<z<z_{max}$ with
$z_{max}$ defined by
\begin{eqnarray}\label{zmax}
 z_{max}=\frac{\pi}{2c}\sqrt{\frac{1}{a(a+b)}}.
\end{eqnarray}
Inverting Eq. (\ref{z_y}), $y(z)$ can be solved easily as
\begin{eqnarray}\label{yz}
y(z)=\frac{1}{c}~\textrm{arcsinh}
         \left[ \sqrt{\frac{a+b}{a}} \tan \left( \sqrt{a(a+b)}\;c z \right) \right].
\end{eqnarray}
Due to this transformation, $\text{e}^{A}$ and $\phi$ can be
rewritten as functions of $z$:
\begin{eqnarray}\label{eA_z}
 \text{e}^{A(z)} =
           \frac{\sqrt{a}(a+b)\sec^{2} \left( \sqrt{a(a+b)}\;c z \right)}
                { \sqrt{a+(a+b)\tan^{2} \left( \sqrt{a(a+b)}\;c z \right)}},
 \end{eqnarray}
 \begin{eqnarray}\label{phi_z}
 \phi=\pm\sqrt{3} \left[ \arctan \left( \sqrt{\frac{a+b}{a}}
                        \tan \big( \sqrt{a(a+b)}\;c z \big) \right)
              + a c z  \right].~~ 
 \end{eqnarray}

In next subsections, it can be seen that the mass-independent
potentials can be obtained conveniently with the conformally
metric (\ref{conflinee2}).

\subsection{Spin-0 scalar fields}

Firstly, we will investigate localization of real scalar fields on
the AdS brane which obtained in previous section. We start by
considering the action of a real scalar $\Phi$ coupled to itself,
gravity and the domain-wall-forming field $\phi$ \cite{DaviesPRD07}:
\begin{eqnarray}
S_0 = \int d^5 x\sqrt{-g} \bigg[- \frac{1}{2} \; g^{M N}
\partial_M \Phi \partial_N \Phi - V(\Phi, \phi)\bigg],
\label{scalarAction}
\end{eqnarray}
where $V(\Phi, \phi)$ is a coupling potential of $\Phi$ to itself
and to the domain-wall-forming field $\phi$, which also called as
Higgs potential, and $V(\Phi, \phi)$ should include $\Phi$,
$\Phi^{2}$, $\Phi^{3}$, $\Phi^{4}$ and $(\phi\Phi)^{2}$ terms, but
by considering a discrete symmetry, $\Phi$ and $\Phi^{3}$ terms can
be eliminated. So we can set \cite{DaviesPRD07}
\begin{eqnarray}\label{coupling_potetial}
V(\Phi, \phi)=
        (\lambda \phi^{2}-u^{2})\Phi^{2}+\tau\Phi^{4}.
\end{eqnarray}
Using the conformal metric (\ref{conflinee2}), the equation of
motion derived from (\ref{scalarAction}) reads as
\begin{eqnarray}
 \frac{1}{\sqrt{-\hat{g}}}\partial_\mu(\sqrt{-\hat{g}}
   \hat{g}^{\mu \nu}\partial_\nu \Phi)
 +e^{-3A} \partial_z
  \left(e^{3A}\partial_z \Phi \right)
   - e^{2A}U(\phi)\Phi = 0, \label{scalarEOM}
\end{eqnarray}
where $U(\phi)$ is defined by
\begin{eqnarray}\label{define_U}
\frac{\partial V(\Phi, \phi)}{\partial \Phi}= U(\phi)\Phi +
\mathcal{O}(\Phi^{3}),
\end{eqnarray}
and is calculated from Eq. (\ref{coupling_potetial}) as
\begin{eqnarray}
U(\phi)=2(\lambda \phi^{2}-u^{2}).
\end{eqnarray}
 Then, by making use of the KK decomposition $\Phi(x,z) =
\sum_n \Phi_n(x)\chi_n(z)e^{-3A/2}$ and demanding $\Phi_n(x)$
satisfies the 4-dimensional massive Klein--Gordon equation
\begin{eqnarray}\label{KGEq}
 \left(\frac{1}{\sqrt{-\hat{g}}}\partial_\mu(\sqrt{-\hat{g}}
   \hat{g}^{\mu \nu}\partial_\nu) -m_{n}^2 \right)\Phi_n(x)=0,
\end{eqnarray}
we can obtain the equation for the scalar KK mode $\chi_n(z)$:
\begin{eqnarray}
  \left[-\partial^2_z+ V_0(z)\right]{\chi}_n(z)
  =m_{n}^2 {\chi}_n(z),
  \label{SchEqScalar1}
\end{eqnarray}
which is a Schr\"{o}dinger equation with the effective potential
given by
\begin{eqnarray}
  V_0(z)&=&\frac{3}{2} \partial^2_{z}A + \frac{9}{4}(\partial_{z}A)^{2}
   + e^{2A}U(\phi)\nonumber\\
        &=& \frac{3}{2} \partial^2_{z}A + \frac{9}{4}(\partial_{z}A)^{2}
   + 2e^{2A}(\lambda\phi^{2}-u^{2}). \label{VScalar}
\end{eqnarray}
Here $m_n$ is the mass of the KK excitation. It is clear that
$V_0(z)$ defined in (\ref{VScalar}) is a 4-dimensional mass-independent potential.

The full 5-dimensional action (\ref{scalarAction}) reduces to the
standard 4-dimensional action for the massive scalars
\begin{eqnarray}
 S_0=- \frac{1}{2} \sum_{n}\int d^4 x \sqrt{-\hat{g}}
     \bigg(\hat{g}^{\mu\nu}\partial_\mu\phi_{n}
           \partial_\nu\phi_{n}
           +m_{n}^2 \Phi^2_{n}
     \bigg), \label{ScalarEffectiveAction}
\end{eqnarray}
when integrated over the extra dimension, in which it is required
that Eq. (\ref{SchEqScalar1}) is satisfied and the following
orthonormality condition is obeyed:
\begin{eqnarray}
 \int^{z_{max}}_{-z_{max}} dz
 \;\chi_m(z)\chi_n(z)=\delta_{mn}.
 \label{normalizationCondition1}
\end{eqnarray}

We will consider two cases in this section. Firstly, we choose
$U(\phi)=0$, that means the scalar field $\Phi$ does not couple to
itself and the background scalar field $\phi(z)$. Secondly, we set
$U(\phi)= 2(\lambda \phi^{2}+u^{2})$, and consider the scalar field
$\Phi$ coupled to itself and $\phi(z)$.

\begin{figure}[htb]
\begin{center}
\includegraphics[width=7cm]{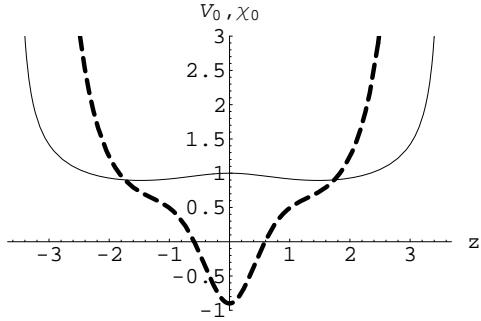}
\end{center}
\caption{The shapes of the potential $V_{0}(z)$ (the dashed line) and
the zero mode $\chi_{0}(z)$ (the thin line). The parameters are set as
$\lambda=0$, $a=0.2$, $b=0.8$, $r=1$, $\gamma=-3$ and $u=0$.  }
 \label{fig_Scalar}
\end{figure}

For the AdS brane world solution (\ref{eA_z}) and $U(\phi)=0$,
the potential corresponding to (\ref{VScalar}) is
\begin{eqnarray}\label{V0z}
  V_0(z)&=& -
         \frac{3ac^{2}(a+b)\sec^{2}\left( \sqrt{a(a+b)}\;c z \right)}
              {8\left[2a+b-b \cos \left(2 \sqrt{a(a+b)}\;c z \right) \right]^{2}} \nonumber \\
      && \times \bigg\{ -28a^{2}-24a b-14b^{2} +(12a^{2}+48ab+17b^{2})
           \cos \left( 2\sqrt{a(a+b)}\;c z \right) \nonumber\\
      &&~~~~~  -2b(4a+3b)\cos \left(4 \sqrt{a(a+b)}
           \;c z \right) +3b^{2}\cos \left( 6\sqrt{a(a+b)}\;c z \right)  \bigg\}.
\end{eqnarray}
It is clear that the potential $V_{0}$ has the asymptotic
behavior: $V_{0}(z\rightarrow \pm z_{max})\rightarrow\infty$. For
this potential, we can obtain the scalar zero mode $\chi_{0}(z)$
by setting $m_{0}=0$:
\begin{eqnarray}
\chi_{0}(z)\propto \text{e}^{\frac{3}{2}A(z)}.
\end{eqnarray}
It can be seen that, when $z\rightarrow \pm z_{max}$, this zero mode
has the asymptotic behavior: $\chi_0 \rightarrow \pm\infty$, which
is physically unacceptable. So it is clear that the zero mode could
not be trapped on the AdS thick brane (see Fig. \ref{fig_Scalar}).

\begin{figure}[htb]
\begin{center}
\subfigure[$n=1$]{\label{fig_scalar_bound_a}
\includegraphics[width=3.5cm]{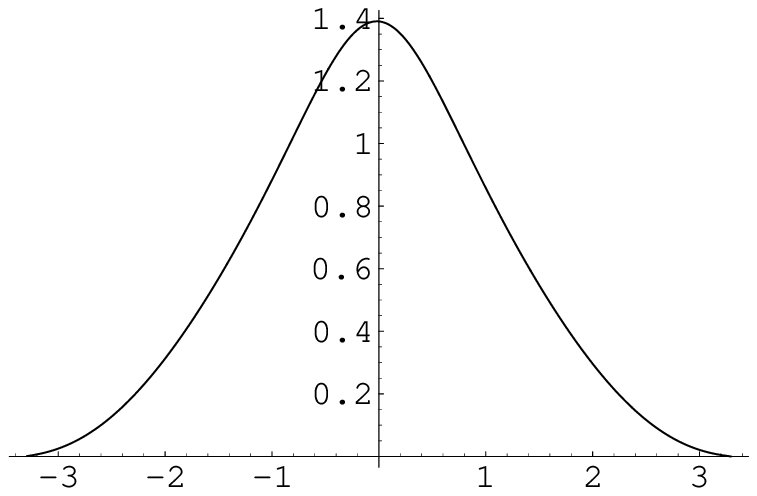}}
\subfigure[$n=2$]{\label{fig_scalar_bound_b}
\includegraphics[width=3.5cm]{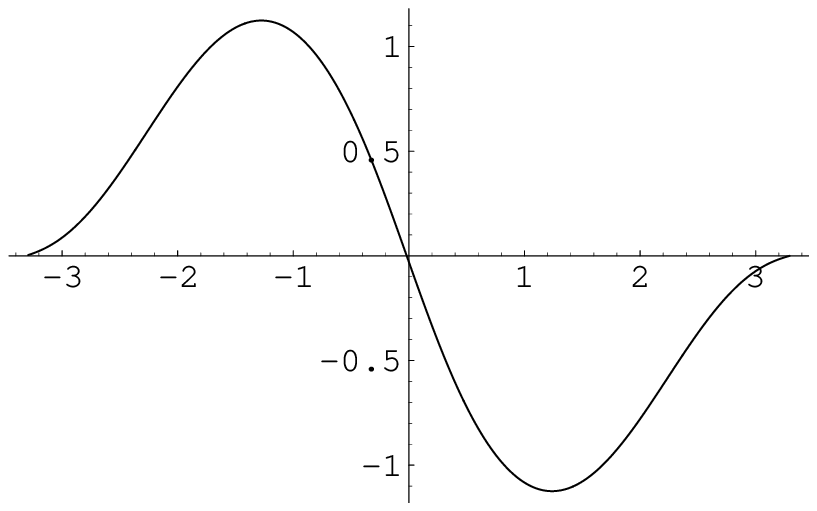}}
\subfigure[$n=3$]{\label{fig_scalar_bound_c}
\includegraphics[width=3.5cm]{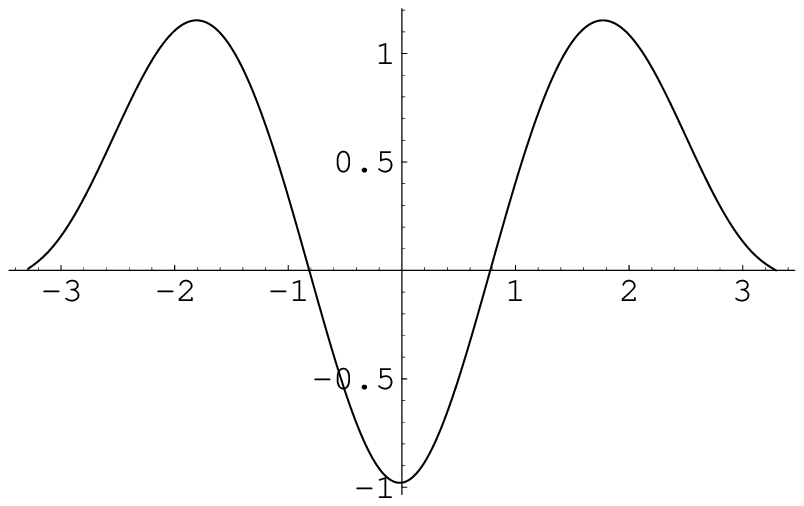}}
\subfigure[$n=4$]{\label{fig_scalar_bound_d}
\includegraphics[width=3.5cm]{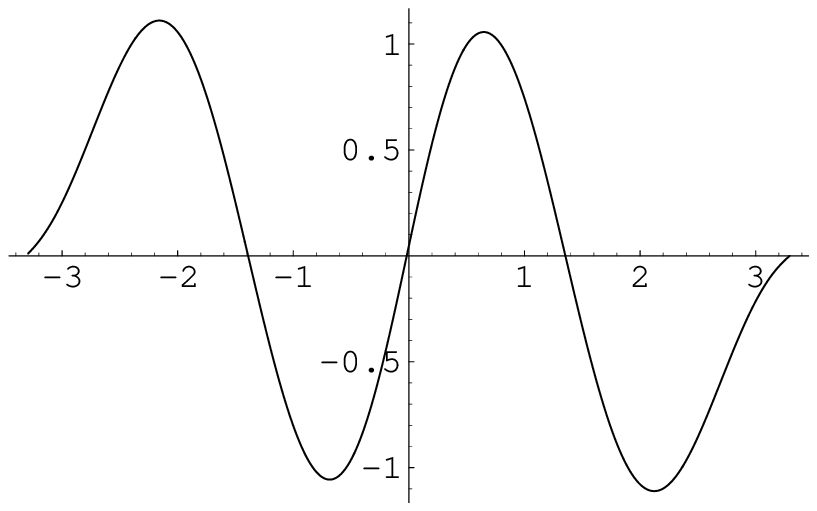}}
\end{center}\vskip -5mm
\caption{The shapes of the scalar bound KK modes $\chi_n(z)$. The
parameters are set as $\lambda=0$, $a=0.2$, $b=0.8$, $r=1$, $\gamma=-3$.}
 \label{fig_scalar_bound}
\end{figure}

\begin{figure}[htb]
\begin{center}
\includegraphics[width=7cm]{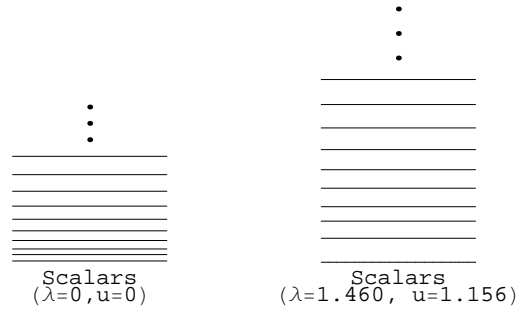}
\end{center}
\caption{The $m_{n}^{2}$ spectrum of the scalar KK modes. The
parameters are set as $a=0.2$, $b=0.8$, $r=1$, $\gamma=-3$.}
\label{Fig_scalar_Spectra}
\end{figure}

However, from the potential $V_0(z)$ (\ref{V0z}), we know that
there are infinite bound KK modes. We can numerically solve Eq.
(\ref{SchEqScalar1}), and the lower bound KK modes are plotted in
Fig. \ref{fig_scalar_bound} and the spectrum of the KK modes is
listed as follows:
\begin{eqnarray}\label{spectra_scalar}
 m_{n}^{2}&=& \{ 0.31, 1.87, 3.25, 5.32, 7.69, 10.52, 13.74, 17.38,  21.44, 25.93, \cdots \},
\end{eqnarray}
where the parameters are set as $a=0.2$, $b=0.8$, $r=1$,
$\gamma=-3$. So, the minimum mass of a 4-dimensional scalar with
mass defined in Eq. (\ref{KGEq}) in background of an AdS brane
cosmology is not zero, and all the 4-dimensional scalars must be
massive and can be localized on the AdS thick brane. We plot the
$m^2_n$ spectrum of the KK modes in Fig. \ref{Fig_scalar_Spectra}.

\begin{figure}[htb]
\begin{center}
\subfigure[$\lambda > \lambda_{0}$]{\label{fig_higgs_scalar_Vz_1}
\includegraphics[width=7cm]{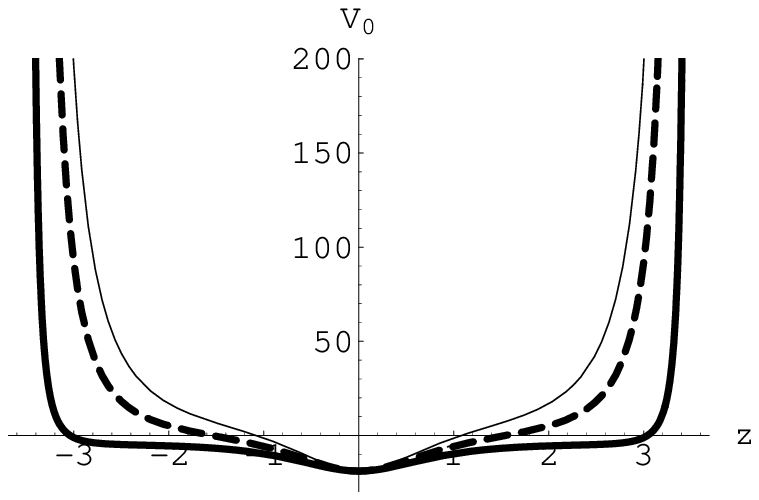}}
\subfigure[$\lambda < \lambda_{0}$]{\label{fig_higgs_scalar_Vz_2}
\includegraphics[width=7cm]{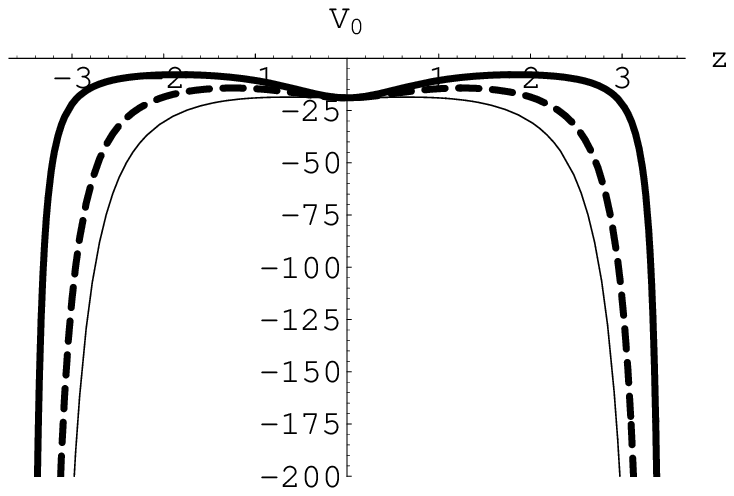}}
\end{center}\vskip -5mm
\caption{The shapes of the potential $V_{0}(z)$. The
parameters are set as $a=0.2$, $b=0.8$, $r=1$, $\gamma=-3$ and $u=3$. In Fig. (a),
we set $\lambda >\lambda_{0}=0.460$, and $\lambda= 0.560$ for the thick line,
$\lambda= 1.460$ for the dashed line, $\lambda= 2.460$ for the thin line.
In Fig. (b), we set $\lambda < \lambda_{0}$, and $\lambda= 0.360$ for the thick line,
$\lambda= -0.540$ for the dashed line, $\lambda= -1.540$ for the thin line.  }
 \label{fig_higgs_scalar_Vz}
\end{figure}

Next, we will consider the case $U(\phi)=2(\lambda \phi^{2}-u^{2})$,
namely, the scalar field $\Phi$ coupled to itself
and the background scalar $\phi(z)$. The potential corresponding to
(\ref{VScalar}) is
\begin{eqnarray}\label{V0z2}
  V_0(z)&=& -
         \frac{3ac^{2}(a+b)\sec^{2}\left( \sqrt{a(a+b)}\;c z \right)}
              {8\left[2a+b-b \cos \left(2 \sqrt{a(a+b)}\;c z \right) \right]^{2}} \nonumber \\
      & & \times \bigg\{ -28a^{2}-24a b-14b^{2} +(12a^{2}+48ab+17b^{2})
           \cos \left( 2\sqrt{a(a+b)}\;c z \right) \nonumber\\
      &&~~  -2b(4a+3b)\cos \left(4 \sqrt{a(a+b)}
           \;c z \right) +3b^{2}\cos \left( 6\sqrt{a(a+b)}\;c z \right)  \bigg\} \nonumber\\
      &&  + \frac{2a(a+b)^{2}\left\{3\lambda
         \left[a c z+ \arctan\left(\sqrt{\frac{a+b}{a}}\tan(\sqrt{a(a+b)}cz)\right) \right]^{2}
          -u^{2} \right\}}
        {\left[a+(a+b)\tan^{2}(\sqrt{a(a+b)}cz)\right]\cos^{4}(\sqrt{a(a+b)}cz)}  .
\end{eqnarray}
When $z\rightarrow \pm z_{max}$, the even potential $V_{0}(z)$ can be reduced to
\begin{eqnarray}
\label{V0z_reduce}
 V_{0}(z\rightarrow \pm z_{max})\rightarrow
  \frac{15 c^2 + 8 \left[\lambda \phi(z_{max})^2 - u^{2}\right]}
       {4c^{2}\left(z \mp \frac{\pi}{2\sqrt{a(a+b)}\;c}\right)^{2}},
\end{eqnarray}
from which we find that the coupling constant $\lambda$ has a critical value
$\lambda_{0}$, which can be expressed as
\begin{eqnarray}\label{lambda_0}
 \lambda_{0}= \frac{8u^{2}-15c^{2}}{8\phi(z_{max})^{2}}.
\end{eqnarray}
For $\lambda > \lambda_{0}$ and $\lambda < \lambda_{0}$, the
potential $V_{0}$ has different asymptotic behavior at
$z\rightarrow\pm z_{max}$. This can be seen from Fig. \ref{fig_higgs_scalar_Vz}.
For $\lambda > \lambda_{0}$,
the potential $V_{0}(z\rightarrow\pm z_{max})\rightarrow +\infty$.
While for $\lambda < \lambda_{0}$,
$V_{0}(z\rightarrow\pm z_{max})\rightarrow -\infty$.
Following, we will discuss the two cases of $\lambda > \lambda_{0}$
and $\lambda < \lambda_{0}$, respectively.

\begin{figure}[htb]
\begin{center}
\subfigure[$n=1$]{\label{fig_scalar_bound2_a}
\includegraphics[width=3.5cm]{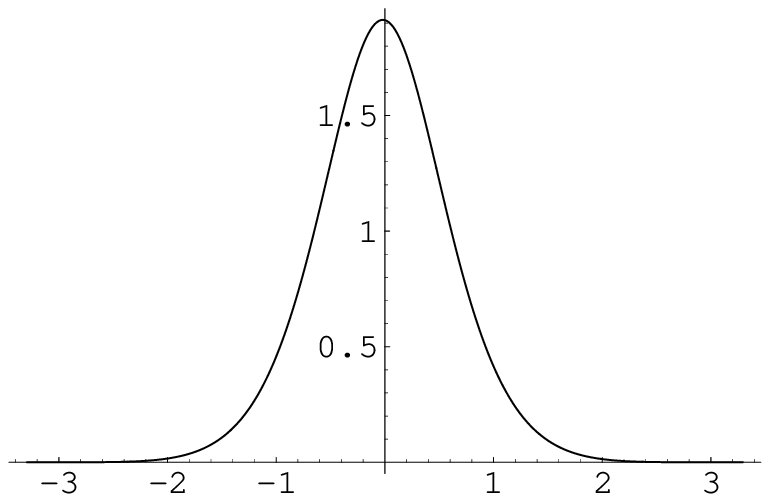}}
\subfigure[$n=2$]{\label{fig_scalar_bound2_b}
\includegraphics[width=3.5cm]{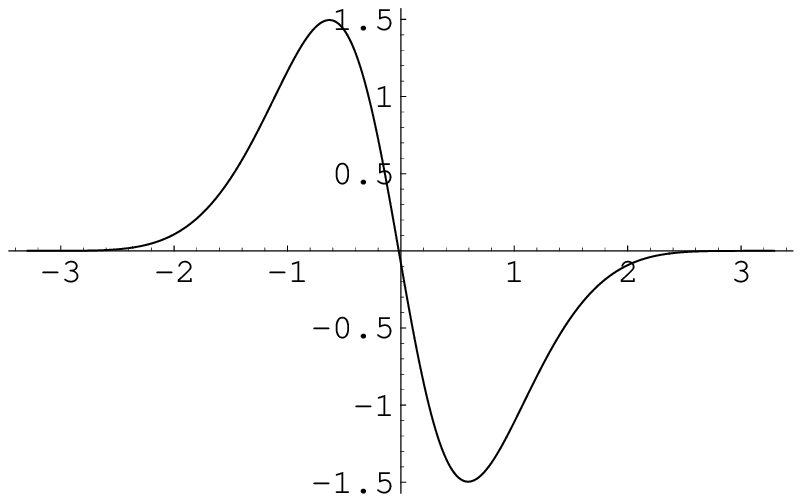}}
\subfigure[$n=3$]{\label{fig_scalar_bound2_c}
\includegraphics[width=3.5cm]{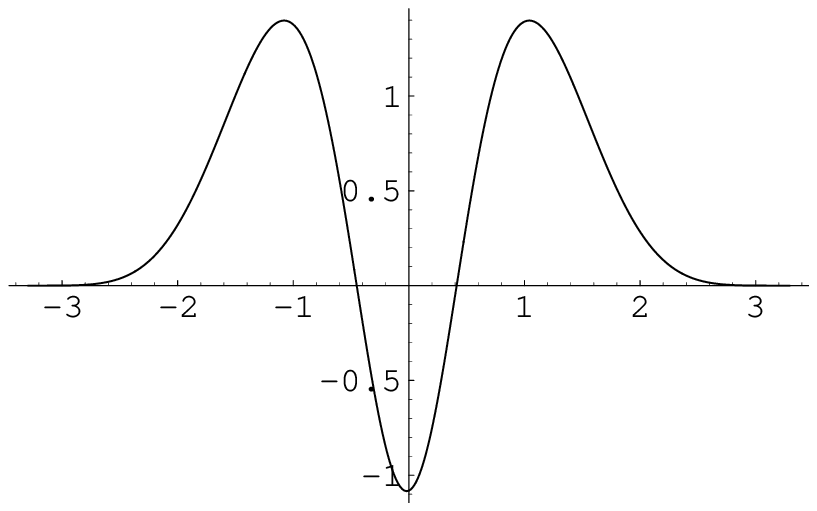}}
\subfigure[$n=4$]{\label{fig_scalar_bound2_d}
\includegraphics[width=3.5cm]{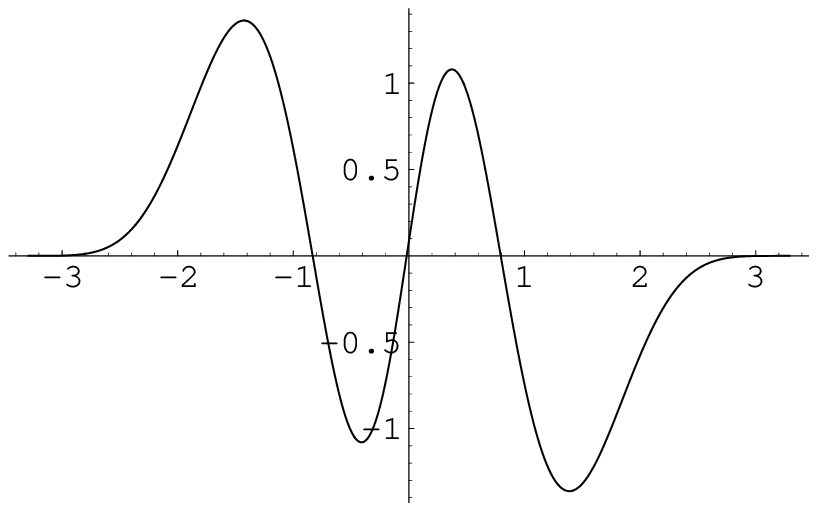}}
\end{center}\vskip -5mm
\caption{The shapes of the scalar bound KK modes $\chi_n(z)$. The
parameters are set as $a=0.2$, $b=0.8$, $r=1$, $\gamma=-3$, $\lambda=1.460$ and $u=1.156$~.}
 \label{fig_scalar_bound2}
\end{figure}

We first consider the case of $\lambda > \lambda_{0}$, for which the potential
$V_{0}$ (\ref{V0z2}) has an infinite deep well (see Fig. \ref{fig_higgs_scalar_Vz_1}),
and supports infinite discrete bound KK modes. When we set $\lambda > \lambda_{0}$ and fix the value of $\lambda$,
if we choose a proper parameter $u$, the zero mode can be localized on the AdS brane,
and if we increase the value of $u$, there will be exist the bound state solutions with $m_{n}^{2}<0$.
In the following discussion, we will set the $\lambda > \lambda_{0}$ and fix the value of $\lambda$,
and choose a proper $u$ for the zero mode localization. Then we can solve numerically Eq.
(\ref{SchEqScalar1}), and the lower bound KK modes are plotted in
Fig. \ref{fig_scalar_bound2} and the spectrum of the KK modes is
listed as follows:
\begin{eqnarray}\label{spectra_scalar_2}
 m_{n}^{2}=\{ 0,5.867, 10.058, 13.593, 18.122, 22.651, 27.568, 32.883, 38.602, 44.726, \cdots \},
\end{eqnarray}
where the parameters are set as $a=0.2$, $b=0.8$, $r=1$,
$\gamma=-3$, $\lambda=1.460$ and $u=1.156$. So from the numerical
solution, we know that the zero mode could be trapped on the AdS
brane by fine-tuning of parameters, and all the 4-dimensional
massive scalar fields can be trapped on the AdS thick brane. We plot
the $m_{n}^{2}$ spectrum of the KK modes in Fig.
\ref{Fig_scalar_Spectra}. Comparing these two cases, we can see that
the four-dimensional scalars would get larger mass when a coupling
potential with a positive coupling constant is introduced.

For the case of $\lambda < \lambda_{0}$, the potential
$V_{0}$ (\ref{V0z2}) has no well (see Fig. \ref{fig_higgs_scalar_Vz_2}),
therefor all the scalar KK modes could not be localized on
the AdS thick brane.

\begin{figure}[htb]
\begin{center}
\subfigure[$\lambda > \lambda_{m}$]{\label{fig_V_Phi_phiz_1}
\includegraphics[width=4.5cm]{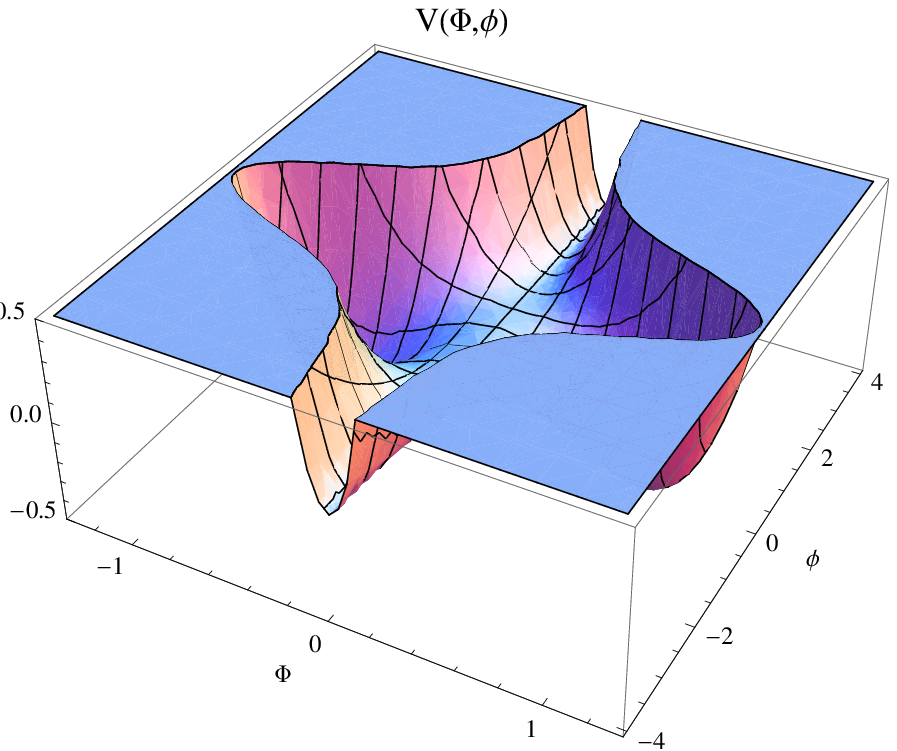}}
\subfigure[$\lambda = \lambda_{m}$]{\label{fig_V_Phi_phiz_2}
\includegraphics[width=4.5cm]{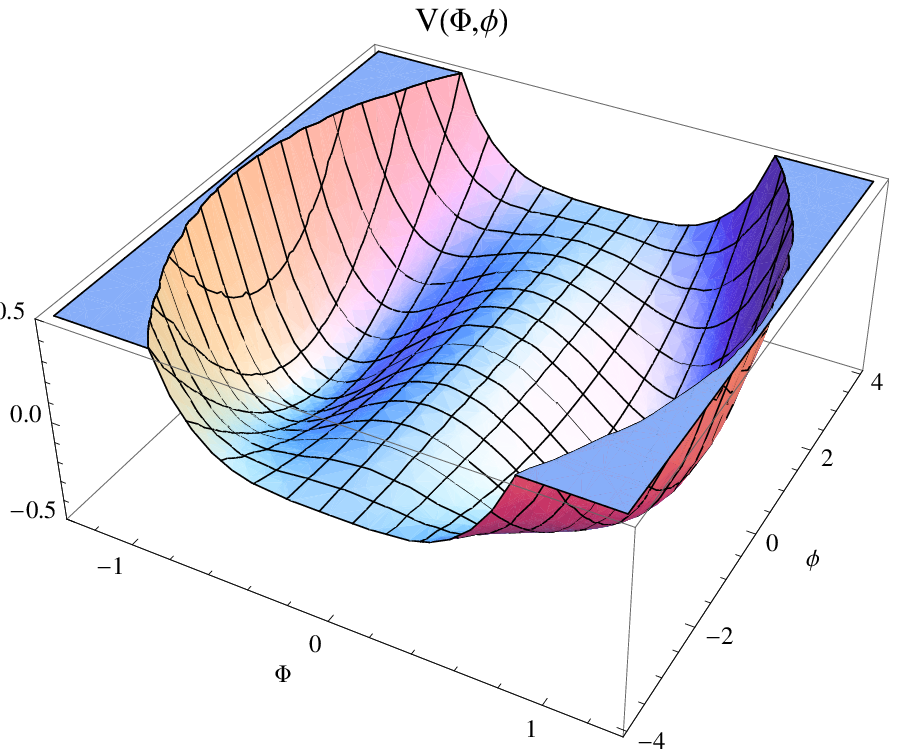}}
\subfigure[$\lambda < \lambda_{m}$]{\label{fig_V_Phi_phiz_3}
\includegraphics[width=4.5cm]{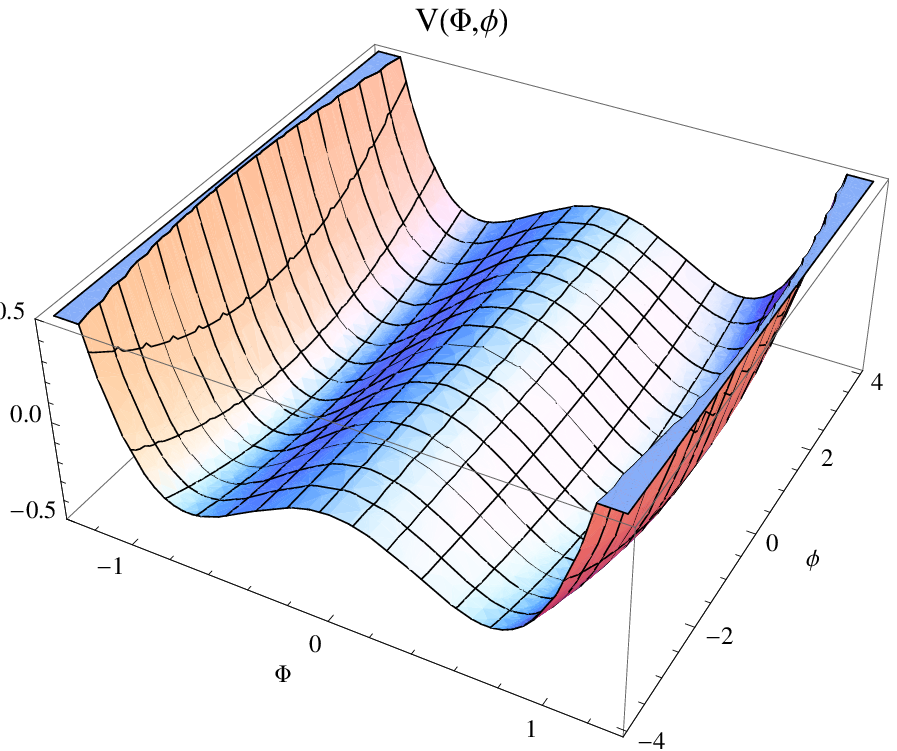}}
\end{center}\vskip -5mm
\caption{The shapes of the potential $V(\Phi,\phi)$.
 The parameters are set as $a=0.2$, $b=0.8$, $r=1$, $\gamma=-3$, $\tau=1$ and $u=1.156$,
 from which we have $\lambda_{m}=0.086$.
 The range of $\phi$ is from $-\phi(z_{max})$ to $+\phi(z_{max})$.
 In Fig. (a), we set $\lambda= 1.460 >\lambda_{m}$.
 In Fig. (b), we set $\lambda=0.086=\lambda_{m}$.
 In Fig. (c), we set $\lambda=0.020<\lambda_{m}$.  }
 \label{fig_V_Phi_phiz}
\end{figure}

At the end of this section we will investigate the property of the
coupling potential. From the expression (\ref{coupling_potetial}),
we can set $\lambda_{m} \phi(z_{max})^{2}-u^{2}=0$, and we can
obtain
\begin{eqnarray}\label{lambda_m}
\lambda_{m}= \frac{u^{2}}{\phi(z_{max})^{2}}.
\end{eqnarray}
When $\lambda > \lambda_{m}$, we have plotted the coupling potential
$V(\Phi, \phi)$ in Fig. \ref{fig_V_Phi_phiz_1}. We can see that, if
$\Phi\rightarrow 0$ as $z\rightarrow\pm z_{max}$,
then $(\phi\;, \Phi)=(\pm\phi(z_{max}), 0)$ are the global minima of the
potential. Because $\phi$ forms the domain wall and $\phi\sim 0$
inside the wall, so the leading term of $V(\Phi, \phi)$
is $-u^{2}\Phi^{2}$, which indicates that the $\Phi=0$
solution is unstable there \cite{DaviesPRD07}. When far from the domain wall,
$\phi\rightarrow \pm\phi(z_{max})$ and
$(\phi\;, \Phi)=(\pm\phi(z_{max}), 0)$ are the global minima of the
potential, so the $\Phi=0$ solution is stable there.
When $\lambda =\lambda_{m}$ (see Fig. \ref{fig_V_Phi_phiz_2}),
the conclusion is the same as the case of $\lambda > \lambda_{m}$.
When $\lambda <\lambda_{m}$ (see Fig. \ref{fig_V_Phi_phiz_3}), there is no stable solution.
From Eqs. (\ref{lambda_0}) and (\ref{lambda_m}), we have $\lambda_m>\lambda_0$,
so when $\lambda>\lambda_m(>\lambda_0)$, the $\Phi=0$ solution is stable at $z\rightarrow\pm z_{max}$
and all the 4-dimensional scalars are localized on the thick brane.

\subsection{Spin-1 vector fields}

\begin{figure}[htb]
\begin{center}
\includegraphics[width=7cm]{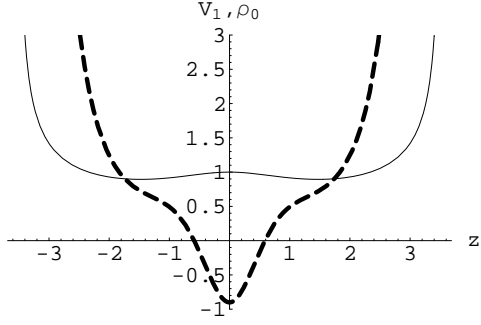}
\end{center}
\caption{The shapes of the potential $V_{1}(z)$ (the dashed line) and
the zero mode $\rho_{0}(z)$ (the thin line). The parameters are set as
$a=0.2$, $b=0.8$, $r=1$, $\gamma=-3$. }
 \label{fig_Vector}
\end{figure}

\begin{figure}[htb]
\begin{center}
\subfigure[$n=1$]{\label{fig_vector_bound_a}
\includegraphics[width=3.5cm]{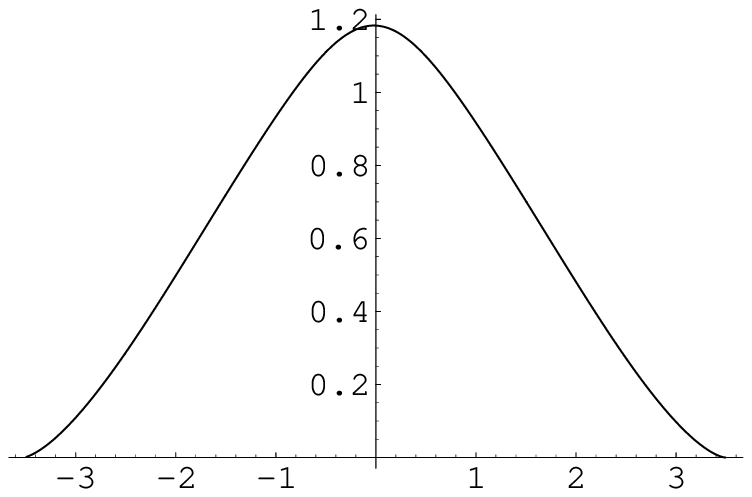}}
\subfigure[$n=2$]{\label{fig_vector_bound_b}
\includegraphics[width=3.5cm]{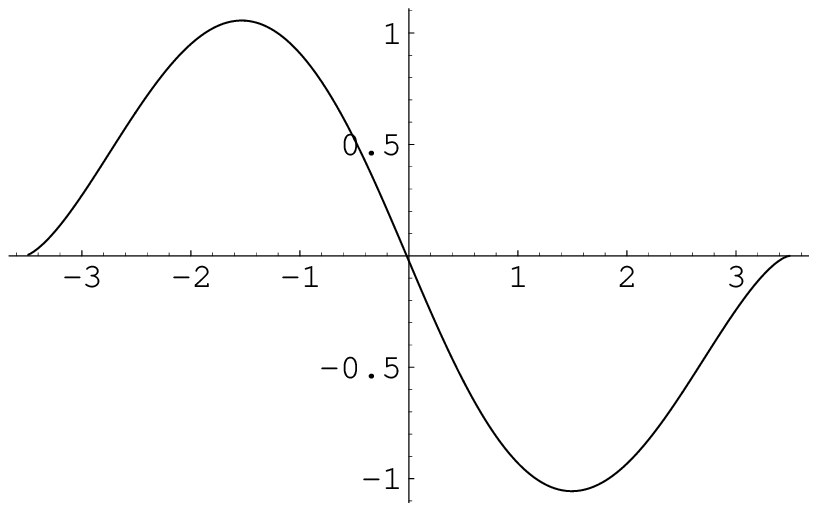}}
\subfigure[$n=3$]{\label{fig_vector_bound_c}
\includegraphics[width=3.5cm]{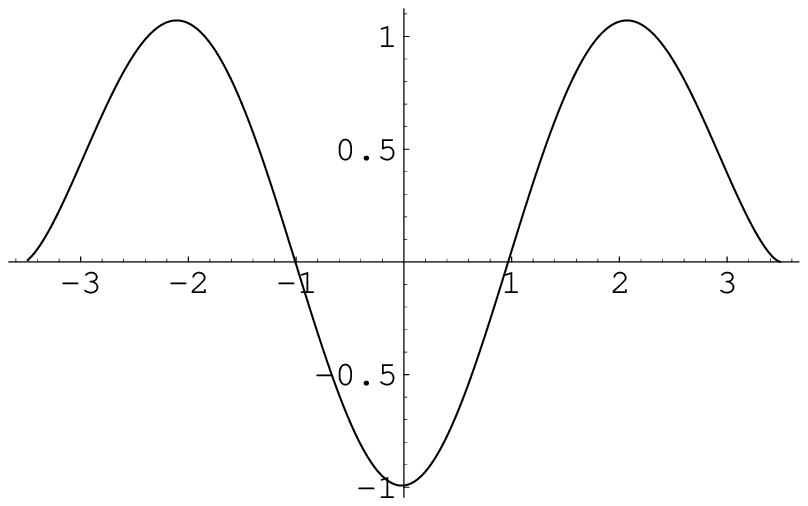}}
\subfigure[$n=4$]{\label{fig_vector_bound_d}
\includegraphics[width=3.5cm]{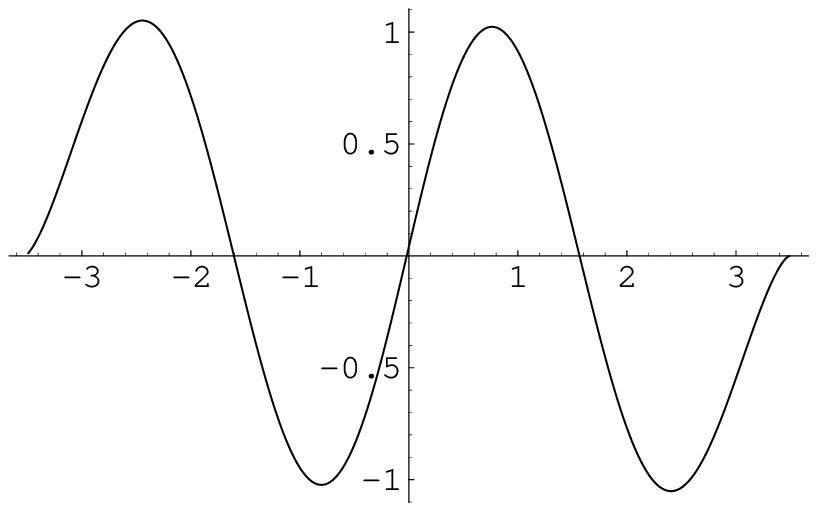}}
\end{center}\vskip -5mm
\caption{The shapes of the bound KK modes $\rho_n(z)$ of vector
fields. The parameters are set as $a=0.2$, $b=0.8$, $r=1$,
$\gamma=-3$. }
 \label{fig_vector_bound}
\end{figure}

\begin{figure}[htb]
\begin{center}
\includegraphics[width=7cm]{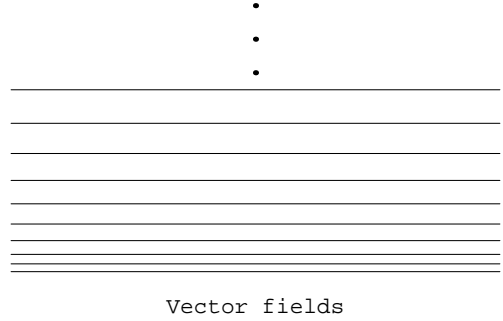}
\end{center}\vskip -5mm
\caption{The spectrum $m^2_n$ of the KK modes of vector fields.
The parameters are set as $a=0.2$, $b=0.8$, $r=1$, $\gamma=-3$.  }
\label{fig_vector_KK}
\end{figure}

Secondly, we will investigate localization of spin-1 vector fields
on the AdS brane. We begin with the 5-dimensional action of a
vector field
\begin{eqnarray}
S_1 = - \frac{1}{4} \int d^5 x \sqrt{-g}~ g^{M N} g^{R S} F_{MR}
F_{NS}, \label{actionVector}
\end{eqnarray}
where $F_{MN} = \partial_M A_N - \partial_N A_M$ is the field tensor
as usual. From this action, one can obtain the equations of motion
\begin{eqnarray}
\frac{1}{\sqrt{-g}} \partial_M (\sqrt{-g} g^{M N} g^{R S} F_{NS}) = 0.
\end{eqnarray}
By using of the background geometry (\ref{conflinee2}), the
equations of motion read as
\begin{eqnarray}
 \frac{1}{\sqrt{-\hat{g}}}\partial_\nu (\sqrt{-\hat{g}} ~
      \hat{g}^{\nu \rho}\hat{g}^{\mu\lambda}F_{\rho\lambda})
    +{\hat{g}^{\mu\lambda}}e^{-A}\partial_z
      \left(e^{A} F_{4\lambda}\right)  = 0, ~~\\
 \partial_\mu (\sqrt{-\hat{g}}~ \hat{g}^{\mu \nu} F_{\nu 4}) =
 0.~~
\end{eqnarray}
Because the fourth component $A_4$ has no zero mode in the effective
4D theory, we assume that it is $Z_2$-odd with respect to the extra
dimension $z$. Furthermore, in order to consistent with the gauge
invariant equation $\oint dz A_4=0$, we choose $A_4=0$ by using
gauge freedom. Under the assumption, the action (\ref{actionVector})
becomes
\begin{eqnarray}
S_1 = - \frac{1}{4} \int d^5 x \sqrt{-g} \bigg\{
        g^{\mu\alpha} g^{\nu\beta} F_{\mu\nu}F_{\alpha\beta}
        +2e^{-A} g^{\mu\nu} \partial_z A_{\mu} \partial_z A_{\nu}
       \bigg\}.
\label{actionVector2}
\end{eqnarray}
Then, using the decomposition of the vector field
$A_{\mu}(x,z)=\sum_n a^{(n)}_\mu(x)\rho_n(z)e^{-A/2}$ and the
orthonormality condition
\begin{eqnarray}
 \int^{z_{max}}_{-z_{max}} dz \;\rho_m(z)\rho_n(z)=\delta_{mn},
 \label{normalizationCondition2}
\end{eqnarray}
the action (\ref{actionVector2}) reduces to
\begin{eqnarray}
S_1 = \sum_{n}\int d^4 x \sqrt{-\hat{g}}~
       \bigg( - \frac{1}{4}\hat{g}^{\mu\alpha} \hat{g}^{\nu\beta}
             f^{(n)}_{\mu\nu}f^{(n)}_{\alpha\beta}
       - \frac{1}{2}m_{n}^2 ~\hat{g}^{\mu\nu}
           a^{(n)}_{\mu}a^{(n)}_{\nu}
       \bigg),
\label{actionVector3}
\end{eqnarray}
where $f^{(n)}_{\mu\nu} = \partial_\mu a^{(n)}_\nu - \partial_\nu
a^{(n)}_\mu$ is the 4-dimensional field strength tensor. In above
reduction, we need that the $\rho_n(z)$ satisfies the following
Schr\"{o}dinger equation
\begin{eqnarray}
  \left[-\partial^2_z +V_1(z) \right]{\rho}_n(z)=m_n^2
  {\rho}_n(z),  \label{SchEqVector1}
\end{eqnarray}
with the mass-independent potential $V$ is given by
\begin{eqnarray}
 V_1(z)&=& -
         \frac{a(a+b)c^{2} \sec^{2}\left( \sqrt{a(a+b)}\;c z \right)}
              {16 \bigg[ a+b\sin^2 \left(\sqrt{a(a+b)}\;c z \right)\bigg]}
               \nonumber \\
   && \times \bigg\{ -20a^{2}-16a b-10b^{2} + (4a^{2}+32ab+11b^{2})
           \cos \left(2 \sqrt{a(a+b)}\;c z \right) \nonumber\\
   &&~~~~ -2b^{2}\cos\left( 4\sqrt{a(a+b)}\;c z \right)+b^{2}\cos \left( 6\sqrt{a(a+b)}\;c z \right)  \bigg\}.
\end{eqnarray}
The potential also has the asymptotic behavior like that of scalar
fields. As same as the discussion about scalar fields, by setting
$m_{0}=0$ we can obtain the zero mode
\begin{eqnarray}
\rho_{0}(z)\propto \text{e}^{\frac{1}{2}A(z)},
\end{eqnarray}
which also could not be trapped on the AdS thick brane (see Fig.
\ref{fig_Vector}). While, it can be seen from the potential $V_{1}$
that there are infinite bound states for the vector KK modes. By
numerical calculation, we can plot the bound states (see Fig.
\ref{fig_vector_bound}) and obtain the mass spectrum:
\begin{eqnarray}\label{spectra_vector}
 m_{n}^{2}= \{0.28, 1.21, 2.34, 3.98, 5.96, 8.37, 11.16, 14.37, 17.97, 21.97, \cdots \},
\end{eqnarray}
where we have set $a=0.2$, $b=0.8$, $r=1$, $\gamma=-3$. The mass
of ground state is $m_{1}=\sqrt{0.28}=0.53$. It is shown that a
spin-1 massless vector field is not localized on AdS thick brane.
However, massive vector fields can be localized on the AdS thick
brane. The mass spectrum (\ref{spectra_vector}) is plotted in Fig.
\ref{fig_vector_KK}.

\subsection{Spin-1/2 fermion fields}

Finally, we will study localization of fermions on the AdS thick
brane. In five dimensions, fermions can be described by four
component spinors and their Dirac structure can be described by
$\Gamma^M= e^M_{~\bar{M}} \Gamma^{\bar{M}}$ with $e^M_{~\bar{M}}$
the vielbein and $\{\Gamma^M,\Gamma^N\}=2g^{MN}$. In this paper,
$\bar{M}, \bar{N}, \cdots =0,1,2,3,5$ and $\bar{\mu}, \bar{\nu},
\cdots =0,1,2,3$ denote the 5D and 4D local Lorentz indices
respectively, and $\Gamma^{\bar{M}}$ are the flat gamma matrices
in five dimensions. In our set-up, the vielbein is given by
\begin{eqnarray}
e_M ^{~~\bar{M}}= \left(%
\begin{array}{ccc}
  \text{e}^{A} \hat{e}_\mu^{~\bar{\nu}} & 0  \\
  0 & \text{e}^{A}  \\
\end{array}%
\right),\label{vielbein_e}
\end{eqnarray}
$\Gamma^M=\text{e}^{-A}(\hat{e}^{\mu}_{~\bar{\nu}}
\gamma^{\bar{\nu}},\gamma^5)=\text{e}^{-A}(\gamma^{\mu},\gamma^5)$,
where $\gamma^{\mu}=\hat{e}^{\mu}_{~\bar{\nu}}\gamma^{\bar{\nu}}$,
$\gamma^{\bar{\nu}}$ and $\gamma^5$ are the usual flat gamma
matrices in the 4D Dirac representation. The Dirac action of a
massless spin-1/2 fermion coupled to gravity and the background
scalar $\phi(y)$ can be expressed as
\begin{eqnarray}
S_{\frac{1}{2}} = \int d^5 x \sqrt{-g} \left[\bar{\Psi} \Gamma^M
          (\partial_M+\omega_M) \Psi
          -\eta \bar{\Psi} F(\phi)\Psi\right]. \label{DiracAction}
\end{eqnarray}
Here $\omega_M$ is the spin connection defined as $\omega_M=
\frac{1}{4} \omega_M^{\bar{M} \bar{N}} \Gamma_{\bar{M}}
\Gamma_{\bar{N}}$ and
\begin{eqnarray}
 \omega_M ^{\bar{M} \bar{N}}
   &=& \frac{1}{2} {e}^{N \bar{M}}(\partial_M e_N^{~\bar{N}}
                      - \partial_N e_M^{~\bar{N}})
    - \frac{1}{2} {e}^{N\bar{N}}(\partial_M e_N^{~\bar{M}}
                      - \partial_N e_M^{~\bar{M}})  \nn \\
   && - \frac{1}{2} {e}^{P \bar{M}} {e}^{Q \bar{N}} (\partial_P e_{Q
{\bar{R}}} - \partial_Q e_{P {\bar{R}}}) {e}_M^{~\bar{R}}.
\end{eqnarray}
The non-vanishing components of the spin connection $\omega_M$ for
the background metric (\ref{conflinee2}) are
\begin{eqnarray}
  \omega_\mu =\frac{1}{2}(\partial_{z}A) \gamma_\mu \gamma_5
             +\hat{\omega}_\mu, \label{spinConnection}
\end{eqnarray}
where $\mu=0,1,2,3$ and $\hat{\omega}_\mu=\frac{1}{4}
\bar\omega_\mu^{\bar{\mu} \bar{\nu}} \Gamma_{\bar{\mu}}
\Gamma_{\bar{\nu}}$ is the spin connection derived from the metric
$\hat{g}_{\mu\nu}(x)=\hat{e}_{\mu}^{~\bar{\mu}}(x)
\hat{e}_{\nu}^{~\bar{\nu}}(x)\eta_{\bar{\mu}\bar{\nu}}$. Then we
can obtain the equation of motion
\begin{eqnarray}
 \left\{ \gamma^{\mu}(\partial_{\mu}+\hat{\omega}_\mu)
         + \gamma^5 \left(\partial_z  +2 \partial_{z} A \right)
         -\eta\; \text{e}^A F(\phi)
 \right \} \Psi =0, \label{DiracEq1}
\end{eqnarray}
where $\gamma^{\mu}(\partial_{\mu}+\hat{\omega}_\mu)$ is the Dirac
operator on the brane.

Now we will investigate the 5-dimensional Dirac equation
(\ref{DiracEq1}), and write the spinor in terms of 4-dimensional
effective fields. On account of the Dirac structure of the fifth
gamma matrix $\gamma^5$, we expect that the left- and right-handed
projections of the four dimensional part to behave differently.
From the equation (\ref{DiracEq1}), we will search for the
solutions of the general chiral decomposition
\begin{equation}
 \Psi= \text{e}^{-2A}\left(\sum_n\psi_{Ln}(x) L_n(z)
 +\sum_n\psi_{Rn}(x) R_n(z)\right),
\end{equation}
where $\psi_{Ln}(x)=-\gamma^5 \psi_{Ln}(x)$ and
$\psi_{Rn}(x)=\gamma^5 \psi_{Rn}(x)$ are the left-handed and
right-handed components of a 4D Dirac field, respectively. Hence,
we assume that $\psi_{L}(x)$ and $\psi_{R}(x)$ satisfy the 4D
massive Dirac equations
$\gamma^{\mu}(\partial_{\mu}+\hat{\omega}_\mu)\psi_{Ln}(x) =
m_n\psi_{R_n}(x)$ and
$\gamma^{\mu}(\partial_{\mu}+\hat{\omega}_\mu)\psi_{Rn}(x) =
m_n\psi_{L_n}(x)$. Then the KK modes $L_n(z)$ and $R_n(z)$ satisfy
the following coupled equations
\begin{subequations}\label{CoupleEq1}
\begin{eqnarray}
 \left[\partial_z
                  + \eta\;\text{e}^A F(\phi) \right]L_n(z)
  &=&  ~~m_n R_n(z), \label{CoupleEq1a}  \\
 \left[\partial_z
                  - \eta\;\text{e}^A F(\phi) \right]R_n(z)
  &=&  - m_n L_n(z). \label{CoupleEq1b}
\end{eqnarray}
\end{subequations}
From the above coupled equations, we can obtain the
Schr\"{o}dinger-like equations for the left- and right-chiral KK
modes of fermions
\begin{eqnarray}
  \big(-\partial^2_z + V_L(z) \big)L_n
            &=&m_{L_n}^{2} L_n,~~
   \label{SchEqLeftFermion}  \\
  \big(-\partial^2_z + V_R(z) \big)R_n
            &=&m_{R_n}^{2} R_n,
   \label{SchEqRightFermion}
\end{eqnarray}
where the mass-independent potentials are given by
\begin{subequations}\label{Vfermion}
\begin{eqnarray}
  V_L(z)&=& \big(\eta\text{e}^{A} F(\phi)\big)^2
     - \partial_z \big(\eta\text{e}^{A} F(\phi)\big), \label{VL}\\
  V_R(z)&=&   V_L(z)|_{\eta \rightarrow -\eta}. \label{VR}
\end{eqnarray}
\end{subequations}

For the purpose of getting the standard 4-dimensional action for the
massive chiral fermions:
\begin{eqnarray}
 S_{\frac{1}{2}} &=& \int d^5 x \sqrt{-g} ~\bar{\Psi}
     \left(  \Gamma^M (\partial_M+\omega_M)
     -\eta F(\phi)\right) \Psi  \nn \\
  &=&\sum_{n}\int d^4 x \sqrt{-\hat{g}}
    ~\bar{\psi}_{n}
      [\gamma^{\mu}(\partial_{\mu}+\hat{\omega}_\mu)
        -m_{n}]\psi_{n},~~~
\end{eqnarray}
we need the following orthonormality conditions for $L_n$ and
$R_n$:
\begin{eqnarray}
 \int_{-z_{max}}^{z_{max}} L_m L_ndz
   &=& \delta_{mn}, \label{orthonormalityFermionL} \\
 \int_{-z_{max}}^{z_{max}} R_m R_ndz
   &=& \delta_{mn}, \label{orthonormalityFermionR}\\
 \int_{-z_{max}}^{z_{max}} L_m R_ndz
   &=& 0. \label{orthonormalityFermionR}
\end{eqnarray}

\begin{figure}[htb]
\begin{center}
\includegraphics[width=7cm]{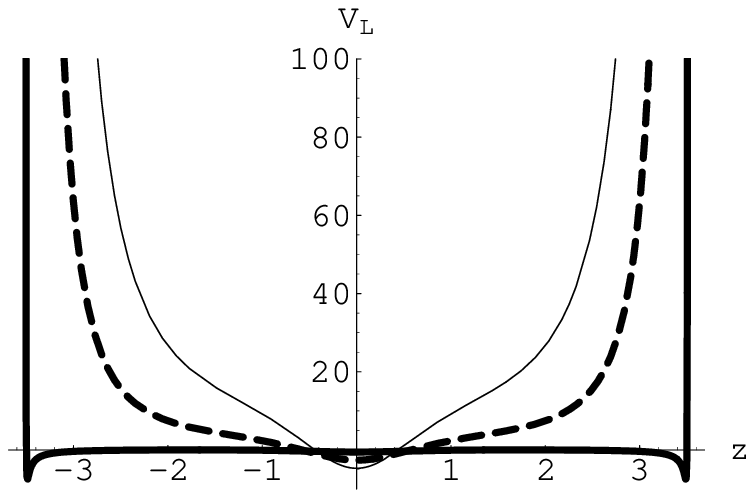}
\includegraphics[width=7cm]{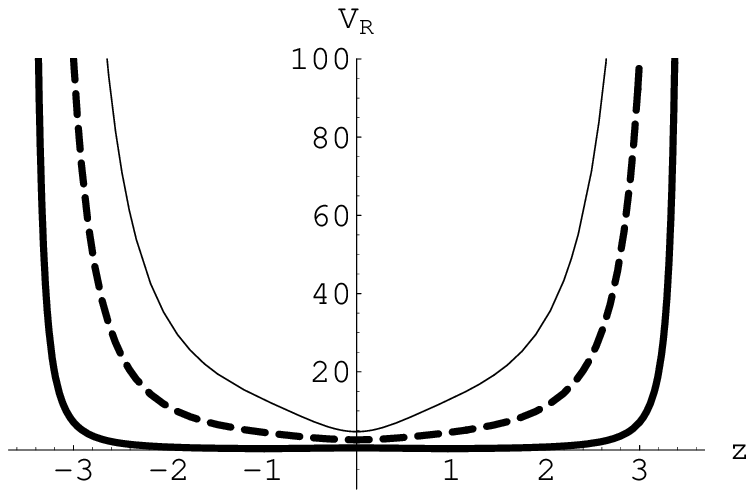}
\end{center}
\caption{The shapes of the potentials $V_{L}(z)$ and $V_{R}(z)$
for $\eta > \eta_0$. The parameters are set as $a=0.2$, $b=0.8$,
$r=1$, $\gamma=-3$, $\eta=0.255$ (the thick line), $\eta=1.254$
(the dashed line), $\eta=2.254$ (the thin line).}
 \label{fig_fermions}
\end{figure}

\begin{figure}[htb]
\begin{center}
\includegraphics[width=7cm]{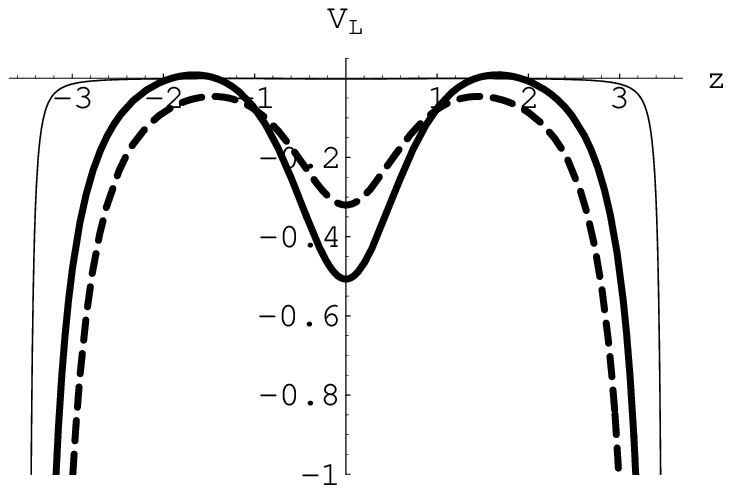}
\includegraphics[width=7cm]{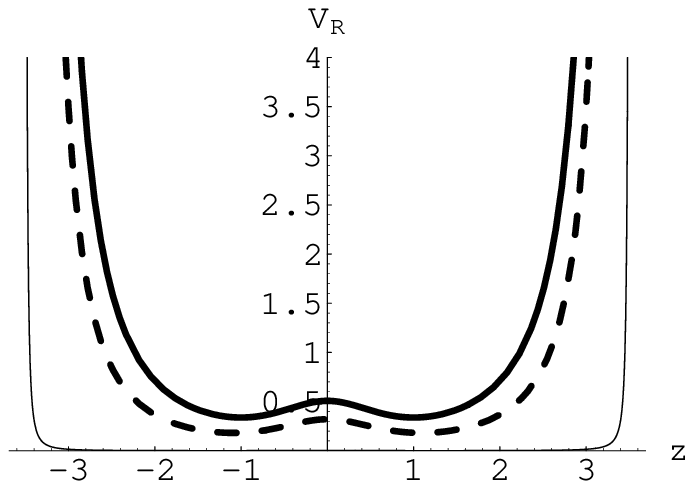}
\end{center}
\caption{The shapes of the potentials $V_{L}(z)$ and $V_{R}(z)$
for $0<\eta < \eta_0$. The parameters are set as $a=0.2$, $b=0.8$,
$r=1$, $\gamma=-3$, $\eta=0.244$ (the thick line), $\eta=0.154$
(the dashed line), $\eta=0.001$ (the thin line).}
 \label{fig_ferm_eta}
\end{figure}

From Eqs. (\ref{SchEqLeftFermion}), (\ref{SchEqRightFermion}) and
(\ref{Vfermion}), we can see that, for the left-chiral
(right-chiral) fermion localization, there must be some kind of
scalar-fermion coupling. This situation is similar to the one in
the RS framework \cite{BajcPLB2000,JackiwPRD1976}. Moreover, if we
demand that  $V_{L}(z)$ and $V_{R}(z)$ is $Z_{2}$-even with
respect to the extra dimension $z$, $F(\phi)$ should be an odd
function of $\phi(z)$. In this paper, we choose the simplest
Yukawa coupling: $F(\phi)=\phi$. Then the form of the potentials
(\ref{Vfermion}) can be expressed as
\begin{subequations}\label{Vfermion_z}
\begin{eqnarray}\label{VL_z}
  V_L(z)&=&  \frac{3\varrho^{4}\eta^{2}\left[a c z +
               \arctan\left(\frac{\varrho}{a}\tan({\varrho}cz)\right)  \right]^{2}}
               {a\cos^{2}({\varrho}cz) \big(a+b\sin^{2}({\varrho}cz)\big)}
     +   \frac{\sqrt{3a}(a+b)\eta\sec^{2}({\varrho}cz)}
            {\sqrt{a+(a+b)\tan^{2}({\varrho}cz)}}
            \nonumber\\
        && \times  \bigg\{ -a c- \frac{{\varrho}c}{a+b\sin^{2}({\varrho}cz)}
                \bigg[ {\varrho} + \left(a c z + \arctan\big[ \frac{\varrho}{a}
                   \tan({\varrho}cz) \big]\right)
                \nonumber\\
        & & ~~~~~~~~~~~~~~~~~~~~~~~~~~~~~~~~~~~~
                 \times  \big(a-b\cos(2{\varrho}cz) \tan({\varrho}cz)
               \big)\bigg]\bigg\}, \label{VL_z} \\
  V_R(z)&=&   V_L(z)|_{\eta \rightarrow -\eta}, \label{VR_z}
\end{eqnarray}
\end{subequations}
where $  \varrho\equiv \sqrt{a(a+b)}$.
The shapes of the potentials $V_{L,R}(z)$ are shown in Fig.
\ref{fig_fermions} and Fig. \ref{fig_ferm_eta} for $\eta > \eta_0=
\frac{2 c }{\sqrt{3}\pi(1+\sqrt{{a}/{(a+b)}}\;)}$ and $0<\eta <
\eta_0$, respectively.

From Fig. \ref{fig_fermions}, it can be seen that, for $\eta >
\eta_0$, the potentials have the asymptotic behavior:
$V_{L,R}\rightarrow + \infty$, when $z\rightarrow \pm z_{max}$.
However, if $0<\eta < \eta_0$, the potentials have different
asymptotic behavior:
 $V_{L}\rightarrow -\infty$ and $V_{R}\rightarrow +\infty$,
when $z\rightarrow \pm z_{max}$ (see Fig. \ref{fig_ferm_eta}).
From Fig. \ref{fig_fermions}, it can be seen that, for $\eta >
\eta_0$, only the potential for left chiral fermions has a
negative value at the location of the brane, i.e., $V_{L}(0)<0$
and $V_{R}(0)>0$, so only the zero mode of left chiral fermions
could be trapped on the AdS thick brane. Fig. \ref{fig_ferm_eta}
shows that there is no zero mode for both chiral fermions in the
case of $0<\eta < \eta_0$.

\begin{figure}[htb]
\begin{center}
\includegraphics[width=7cm]{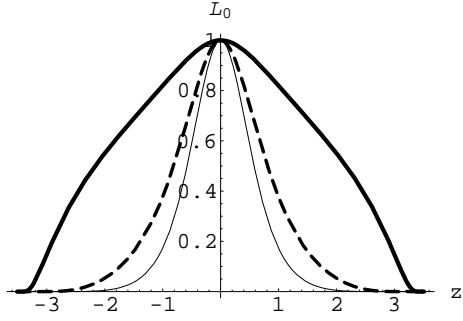}
\end{center}
\caption{The shapes of the zero mode $L_0(z)$ for the case
$\eta>\eta_0$. The parameters are set as $a=0.2$, $b=0.8$, $r=1$,
$\gamma=-3$, $\eta=0.255$ (the thick line), $\eta=1.254$ (the dashed
line), $\eta=2.254$ (the thin line). }
  \label{fig_ferm_zero}
\end{figure}

Setting $\eta > \eta_0$, we can obtain the left chiral fermion zero
mode solved from (\ref{CoupleEq1a}) by setting $m_{0}=0$:
\begin{equation}
 L_0(z)
 \propto \exp\left(-\eta\int^z_0 dz'\text{e}^{A(z')}\phi(z')\right).
  \label{zeroMode1}
\end{equation}
This integral is very complex, so we could not give an analytical
expression. However, we can plot the shape of zero mode by
numerically integral (see Fig. \ref{fig_ferm_zero}).

\begin{figure}[htb]
\begin{center}
\subfigure[$n=0$]{\label{fig_lfet_bound_a}
\includegraphics[width=3.5cm]{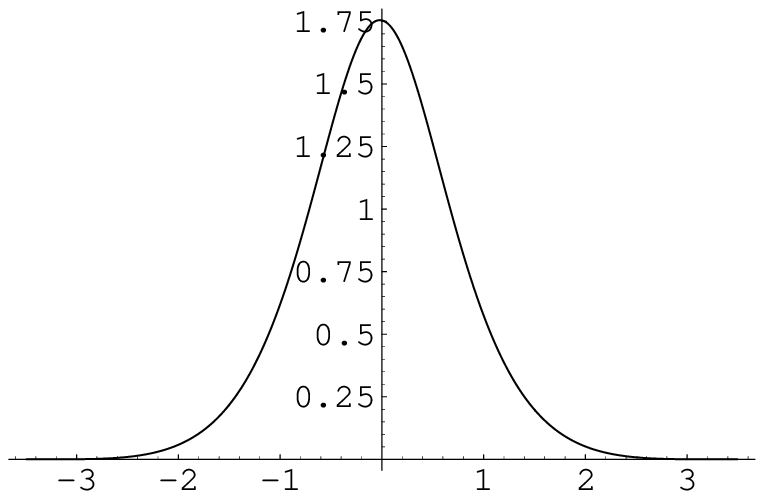}}
\subfigure[$n=1$]{\label{fig_lfet_bound_b}
\includegraphics[width=3.5cm]{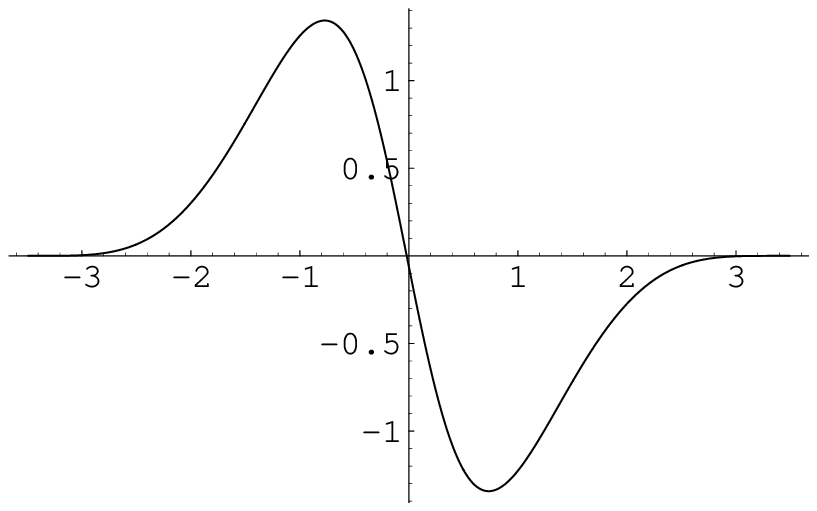}}
\subfigure[$n=2$]{\label{fig_lfet_bound_c}
\includegraphics[width=3.5cm]{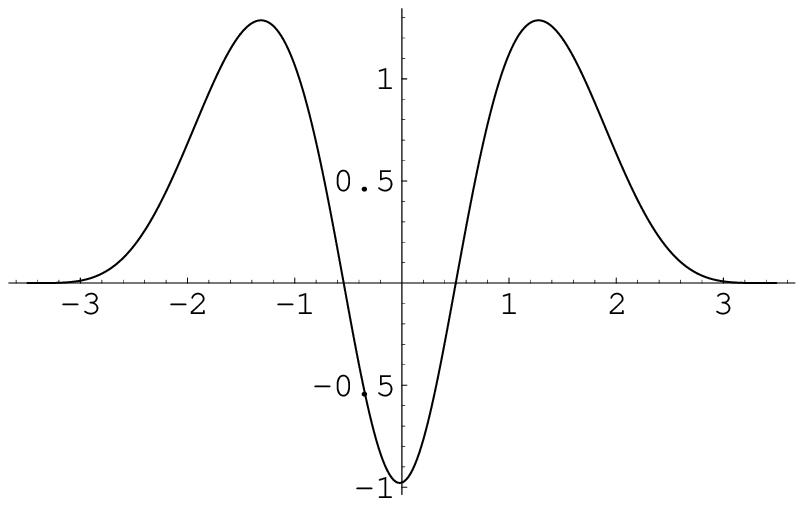}}
\subfigure[$n=3$]{\label{fig_lfet_bound_d}
\includegraphics[width=3.5cm]{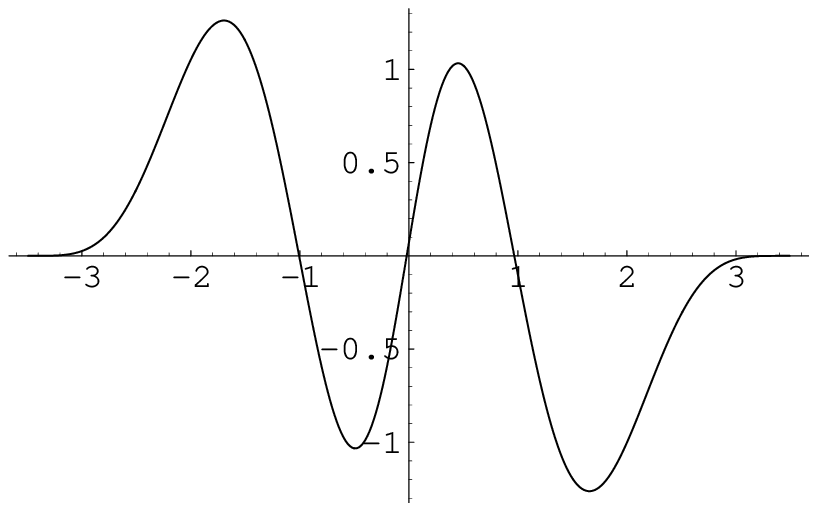}}
\end{center}\vskip -5mm
\caption{The shapes of the left chiral fermions bound states
$L_n(z)$ for the case $\eta>\eta_0$. The parameters are set as
$a=0.2$, $b=0.8$, $r=1$, $\gamma=-3$, $\eta=1.254$.  }
 \label{fig_lfet_bound}
\end{figure}

\begin{figure}[htb]
\begin{center}
\subfigure[$n=1$]{\label{fig_right_bound_a}
\includegraphics[width=3.5cm]{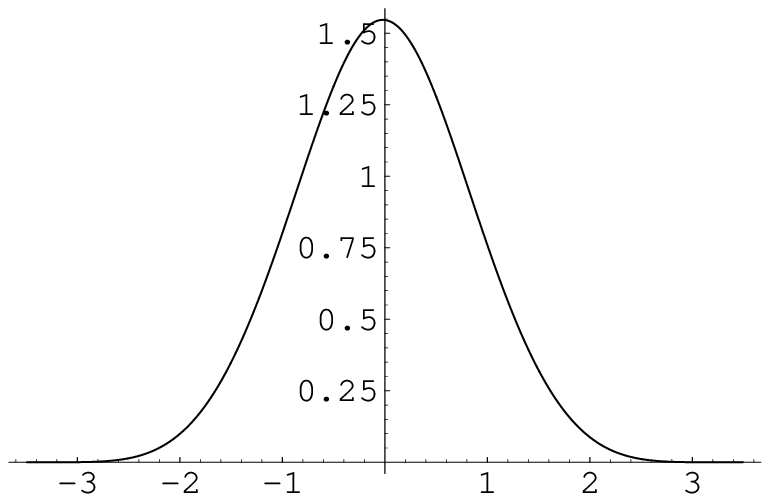}}
\subfigure[$n=2$]{\label{fig_right_bound_b}
\includegraphics[width=3.5cm]{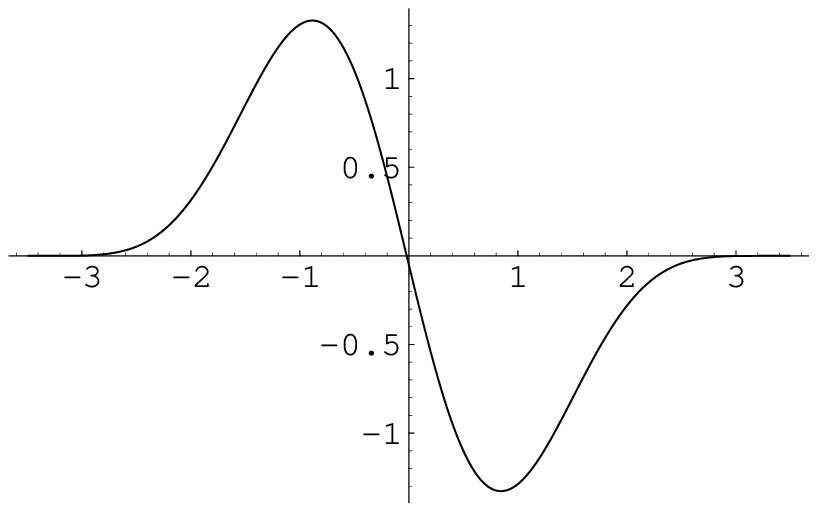}}
\subfigure[$n=3$]{\label{fig_right_bound_c}
\includegraphics[width=3.5cm]{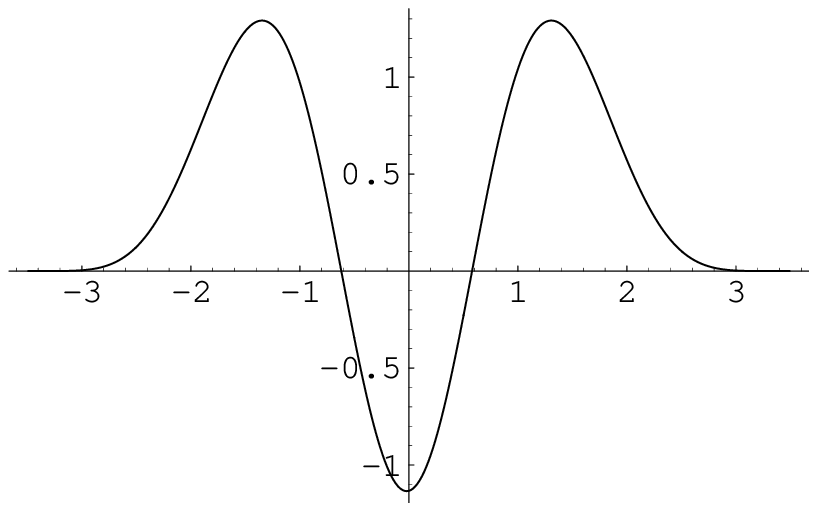}}
\subfigure[$n=4$]{\label{fig_right_bound_d}
\includegraphics[width=3.5cm]{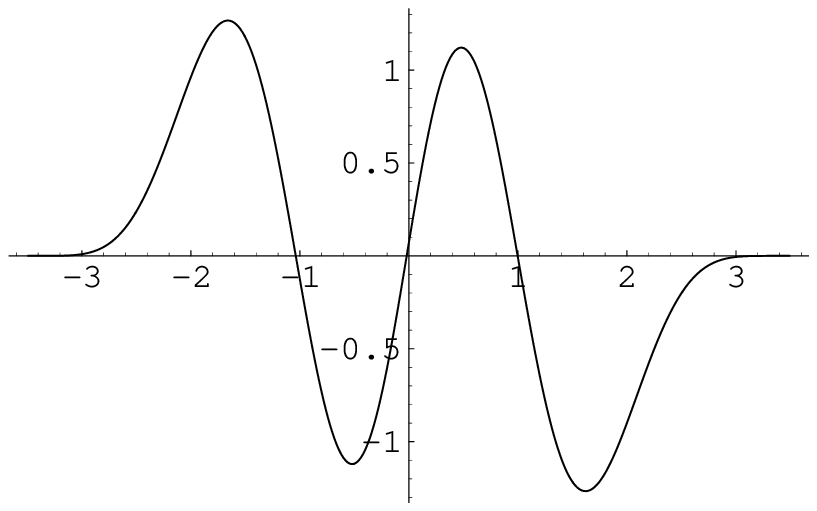}}
\end{center}\vskip -5mm
\caption{The shapes of the right chiral fermions bound states
$R_n(z)$ for the case $\eta>\eta_0$. The parameters are set as
$a=0.2$, $b=0.8$, $r=1$, $\gamma=-3$ $\eta=1.254$.}
 \label{fig_right_bound}
\end{figure}

\begin{figure}[htb]
\begin{center}
\includegraphics[width=7cm]{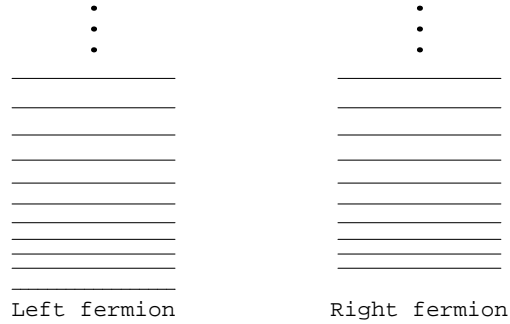}
\end{center}
\caption{The $m^2_n$ spectra of left and right chiral fermions for
the case $\eta>\eta_0$. The parameters are set as $a=0.2$,
$b=0.8$, $r=1$, $\gamma=-3$, $\eta=1.254$. }
\label{fig_fermion_spectra}
\end{figure}

Furthermore, when $\eta > \eta_0$, all the KK states of both left
chiral fermions and right chiral fermions are bound states.
However, when $\eta < \eta_0$, just only the right chiral fermions
have infinite bound states. We first consider the case of $\eta >
\eta_{0}$. The Schr\"{o}dinger-like equations for the left and
right chiral fermions can be solved numerically, and the bound
states are plotted in Fig. \ref{fig_lfet_bound} (left chiral
fermions) and Fig. \ref{fig_right_bound} (right chiral fermions).
The discrete mass spectra for both left and right chiral fermions
are calculated as
\begin{subequations}\label{spectra_fermion}
\begin{eqnarray}
 m_{L_n}^{2}&=& \{ 0, 4.04, 6.77, 9.59, 12.79,
        16.39, 20.40,24.81, 29.64, 34.87, 40.51,  \cdots \},  \\
 m_{R_n}^{2}&=& \{~~~ 4.04, 6.77, 9.59, 12.79,
        16.39, 20.40, 24.81, 29.64, 34.87, 40.51,   \cdots \},
\end{eqnarray}\end{subequations}
where the parameters are set as $a=0.2$, $b=0.8$, $r=1$,
$\gamma=-3$, and $\eta=1.254$. The ground state of left chiral
fermions is the zero mode. However, the ground state of right chiral
fermions is massive. The spectra are also shown in Fig.
\ref{fig_fermion_spectra}.

\begin{figure}[htb]
\begin{center}
\subfigure[$n=1$]{\label{fig_right_bound_eta_a}
\includegraphics[width=3.5cm]{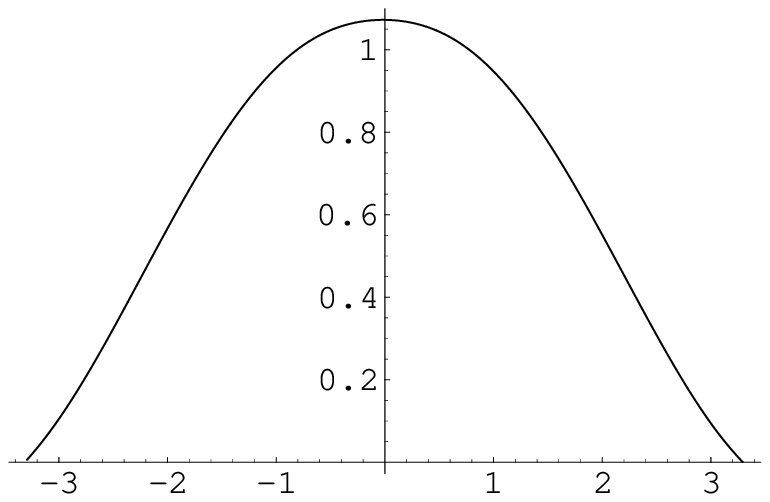}}
\subfigure[$n=2$]{\label{fig_right_bound_eta_b}
\includegraphics[width=3.5cm]{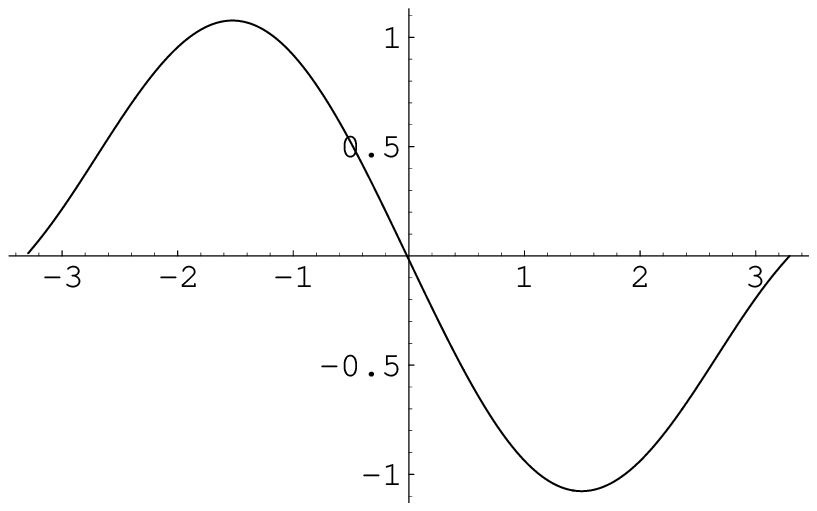}}
\subfigure[$n=3$]{\label{fig_right_bound_eta_c}
\includegraphics[width=3.5cm]{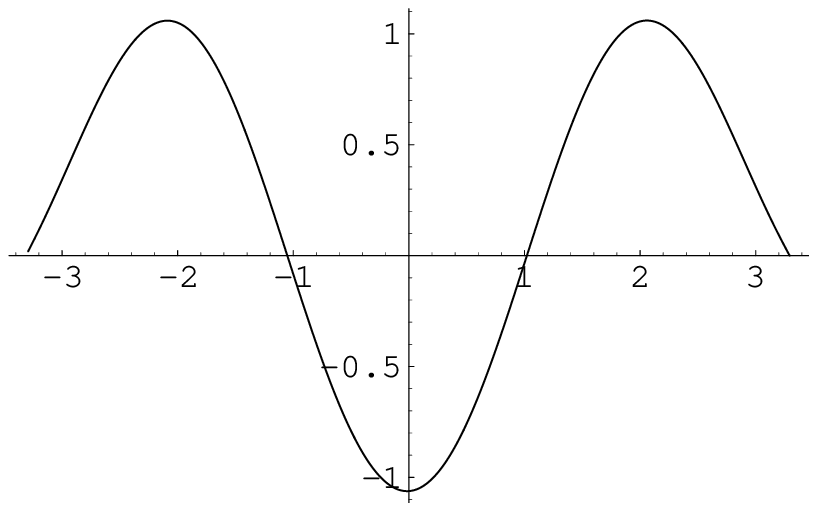}}
\subfigure[$n=4$]{\label{fig_right_bound_eta_d}
\includegraphics[width=3.5cm]{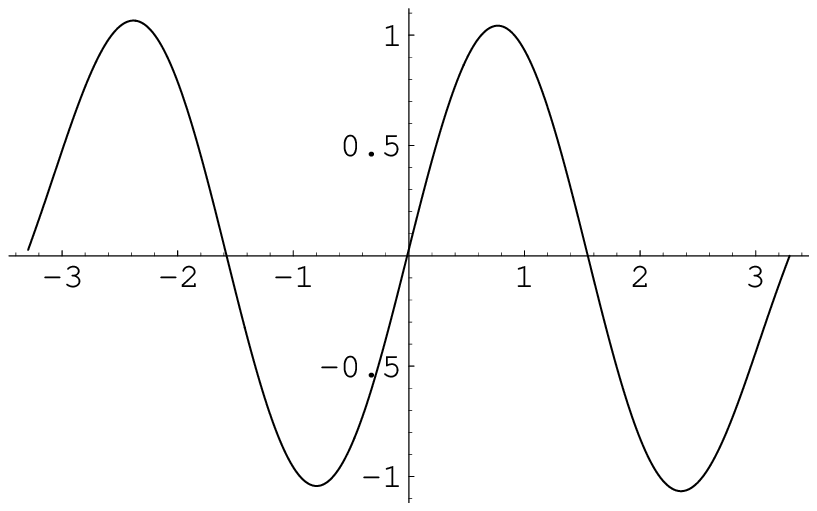}}
\end{center}\vskip -5mm
\caption{The shapes of the right chiral fermions bound states for
the case $0<\eta<\eta_0$. The parameters are set as $a=0.2$,
$b=0.8$, $r=1$, $\gamma=-3$ $\eta=0.154$. }
 \label{fig_right_bound_eta}
\end{figure}

\begin{figure}[htb]
\begin{center}
\includegraphics[width=7cm]{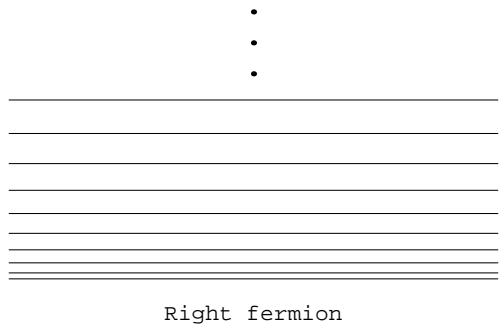}
\end{center}
\caption{The discrete spectra of right chiral fermions mass for
the case $0<\eta<\eta_0$. The parameters are set as $a=0.2$,
$b=0.8$, $r=1$, $\gamma=-3$, $\eta=-0.154$. }
\label{fig_ferm_spectrum_eta}
\end{figure}

Next, we turn to the case $0<\eta < \eta_{0}$. Due to the shape of
the potential $V_L(z)$, left chiral fermions could not be
localized on AdS thick brane. Just only right chiral fermions have
bound states. By numerical method, we can obtain these bound
states, which are plotted in Fig. \ref{fig_right_bound_eta} for
the lower KK states. At the same time, we also get the discrete
mass spectrum of right chiral fermions:
\begin{eqnarray}\label{spectra_fermion_eta}
 m_{R_n}^{2}&=& \{0.51, 1.28, 2.58, 4.26, 6.39, 8.95, 11.95, 15.40, 19.29, 23.62,   \cdots \},
\end{eqnarray}
with the parameters setting as $a=0.2$, $b=0.8$, $c=1$ and
 $\eta= 0.154$. The ground state of right chiral fermions is still
massive. The discrete mass spectrum (\ref{spectra_fermion_eta}) is
shown in Fig. \ref{fig_ferm_spectrum_eta}.

At the end of this section, we make some comments on the issue of
localization of fermions. Localizing fermions on branes requires
us to introduce other interactions besides gravity. In our case,
when coupling coefficient $\eta=0$, both left and right fermions
could not be localized.

\section{Conclusion}
\label{SecConclusion}

In this paper, we have shown the shapes of the mass-independent
potentials of the KK modes of various spin fields in the
corresponding Schr\"{o}dinger equations. In this way, we have
investigated the localization and mass spectra of various matters
with spin-0, 1 and 1/2 on a kind of AdS thick brane.

For spin-0 scalars, if we do not introduce scalar-scalar couplings
$(U=0)$, the situation is same like spin-1 vectors, both potentials
of the KK modes in the corresponding Schr\"{o}dinger equation have a
negative value at the location of the brane, and have the following
asymptotic behavior: $V_{0,1}(z\rightarrow \pm z_{max})\rightarrow +
\infty$, or equivalently, the potentials trend to infinite when far
away from the brane ($y\rightarrow\pm\infty$). Such potentials
suggest that the KK mass spectra of the scalars and vectors are
consisted of infinite discrete bound KK modes. However, the massless
scalar and vector KK modes could not be trapped on the AdS thick
brane. The ground state is massive, and all KK modes are bound
states.

When introducing the scalar $\Phi$ coupling with itself and the
domain-wall-forming field $\phi$ via a Higgs potential for spin-0
scalars, we found that the coupling constant has a critical value
$\lambda_{0}$. We have investigated the cases of $\lambda >
\lambda_{0}$ and $\lambda < \lambda_{0}$, respectively. For $\lambda
> \lambda_{0}$, the result is same as that of $U=0$,
but the massless scalar could be localized by fine-tuning of
parameters. However, for $\lambda < \lambda_{0}$, the potential of
the KK modes in the corresponding Schr\"{o}dinger equation has no
well, and so there is no bound KK mode. We also have investigated
the stability of the solution of $\Phi=0$ inside of the domain wall
and far from the domain wall.

For spin-1/2 fermions, if one does not introduce scalar-fermion
couplings, there is no bound state for both left- and right-hand
KK modes. Hence, we considered the usual Yukawa coupling $\eta
\bar{\Psi} \phi \Psi$, and found that the coupling constant $\eta$
has a critical value
$\eta_{0}=\frac{2c}{\sqrt{3}\pi(1+\sqrt{{a}/{(a+b)}}\;)}$. For
$\eta>\eta_{0}$ and $\eta < \eta_{0}$, the left- and right-hand
KK modes have different properties.

For $\eta > \eta_{0}$, just only the potential for the left-chiral
fermion KK modes has a finite negative well at the location of the brane,
which results in that there exists only the four-dimensional massless left-chiral fermion.
However, both potentials of left- and right-chiral fermion KK modes have
the same asymptotic behavior: $V_{L,R}(z\rightarrow \pm
z_{max})\rightarrow + \infty$. So, both left and right chiral
fermion KK modes have infinite bound states.
Since the pairs of left-hand and right-hand KK modes
couple together through mass terms to become four-dimensional Dirac fermions,
the four-dimensional massive Dirac fermions could be localized on the AdS brane,
and have a set of discrete mass spectrum.

For the situation with $0<\eta<\eta_{0}$, the potential for left-chiral
fermion KK modes is very different from the right one. When
$z\rightarrow \pm a_{max}$, the potentials have different
asymptotic behavior: $V_{L}\rightarrow - \infty$ and
$V_{R}\rightarrow+\infty$. So there is no bound state for
left-chiral fermion KK modes, but the spectrum of right-chiral fermion KK modes is
consisted of infinite discrete bound KK modes. Therefor,
for the case $0<\eta < \eta_{0}$, no four-dimensional Dirac fermion
can be localized on the AdS brane.

\section*{Acknowledgement}

This work was supported by the Program for New Century Excellent
Talents in University, the National Natural Science Foundation of
China (No. 10705013), the Doctoral Program Foundation of
Institutions of Higher Education of China (No. 20070730055 and No. 20090211110028),
the Key Project of Chinese Ministry of Education (No. 109153),
and the Natural Science Foundation of Gansu Province, China (No. 096RJZA055).

\end{document}